\documentclass[conference]{IEEEtran}
\IEEEoverridecommandlockouts
\usepackage{cite}
\usepackage{amsmath,amssymb,amsfonts}
\usepackage{algorithmic}
\usepackage{graphicx}
\usepackage{textcomp}
\usepackage{xcolor}
\def\BibTeX{{\rm B\kern-.05em{\sc i\kern-.025em b}\kern-.08em
    T\kern-.1667em\lower.7ex\hbox{E}\kern-.125emX}}

\usepackage{subcaption}
\PassOptionsToPackage{hyphens}{url}
\usepackage[hyphens]{url}
\usepackage{hyperref}
\usepackage{booktabs}
\usepackage{graphicx}
\usepackage{caption}
\usepackage{pdfpages}
\usepackage{setspace}
\usepackage{subfiles}
\usepackage[font=small,labelfont=bf]{caption}
\usepackage{url}
\usepackage{color}
\usepackage{listings}
\usepackage{xcolor}
\usepackage{setspace}
\usepackage{siunitx}
\usepackage{verbatimbox,lipsum}
\usepackage{fancyvrb}
\usepackage{tcolorbox}
\usepackage{float}
\usepackage{multirow}
\usepackage{amsmath}
\usepackage{amsfonts}
\usepackage{amssymb}
\usepackage{placeins}
\usepackage{mathtools}
\DeclarePairedDelimiter{\ceil}{\lceil}{\rceil}

\usepackage{enumitem}

\newcommand{\name}{Archapt}

\begin{document}

\title{\Large \bf Architecture-Aware, High Performance Transaction \\ for Persistent Memory 
}

\newcommand{\email}[1]{\texttt{\small{#1}}}
\author{
\begin{minipage}{5.0cm}
    \centering
    Kai Wu \\
    \email{kwu42@ucmerced.edu}
  \end{minipage}
  \begin{minipage}{6cm}
    \centering
   Jie Ren\\
    \email{jren6@ucmerced.edu}
  \end{minipage}
  \begin{minipage}{5.0cm}
   \centering
    Dong Li \\
    \email{dli35@ucmerced.edu}
 \end{minipage}  \\\\
\phantom{hahah}University of California, Merced}



\maketitle
\thispagestyle{plain}
\pagestyle{plain}

\begin{abstract}
Byte-addressable non-volatile main memory (NVM) demands transactional mechanisms to access and manipulate data on NVM atomically. Those transaction mechanisms often employ a logging mechanism (undo logging or redo logging). However, the logging mechanisms can bring large runtime overhead (8\%-49\% in our evaluation), and 41\%-78\% of the overhead attributes to the frequent cache-line flushing. Such large overhead significantly diminishes the performance benefits offered by NVM. In this paper, we introduce a new method to reduce the overhead of cache-line flushing for logging-based transactions. Different from the traditional method that works at the program level and leverages program semantics to reduce the logging overhead, we introduce architecture awareness. In particular, we do not flush certain cache blocks, as long as they are estimated to be eliminated out of the cache because of the hardware caching mechanism (e.g., the cache replacement algorithm). Furthermore,
we coalesce those cache blocks with low dirtiness to improve the efficiency of cache-line flushing.
 We implement an architecture-aware, high performance transaction runtime system for persistent memory, \name{}.
Our results show that comparing with an undo logging (PMDK) and a redo logging (Mnemosyne), \name{} reduces cache-line flushing by 66\% and improves system throughput by 19\% on average (42\% at most). 
Our crash tests with four hardware caching policies show that \name{} provides strong a guarantee on crash consistency.
\end{abstract}


\section{Introduction}


Non-volatile memories (NVM), addressed at a byte granularity directly by CPU and accessed roughly at the latency of main memory, are coming. 
While NVM \textit{as main memory} provides an appealing interface that uses simple \texttt{load/store}, it brings 
new challenges to the designs of persistent data structures, storage systems, and databases.
In particular, a \texttt{store} does not immediately make data persistent, because the memory hierarchy (e.g., caches and store buffers) 
can remain non-persistent. There is a need to ensure that data is modified atomically when moving from one consistent state to another, in order to provide \textit{consistency} after a crash due to power loss or hardware failure. 

The NVM challenges have resulted in investigations of transactional mechanisms to access and manipulate data on persistent memory (NVM) atomically~\cite{atlas:oopsla14,Kolli:ASPLOS2016,mnemosyne_asplos11,lu:iccd14,Atomic_ucsd:eurosys17,nvm_t:nvmsa18,nvm_t:Oracle,nvm_t:nvmsa17}. Those transactional mechanisms often employ a logging technique (undo logging or redo logging). 
However, those transactional mechanisms have a high overhead. Our performance evaluation reveals that 
running TPC-C~\cite{TPCC,pytpcc} and YCSB (A-F)~\cite{YCSB,YCSB:github} against Redis~\cite{Redis}, and OLTP-bench~\cite{oltpbench:github} (TPC-C, LinkBench~\cite{linkbench:simod13} and YCSB) against SQLite~\cite{sqlite}  based on an implementation of undo logging from Intel PMDK~\cite{pmdk}) or a redo logging from~\cite{mnemosyne_asplos11} to build transactions, we have overheads of 8\%-49\%. Such large overhead significantly diminishes the performance benefit NVM promises to provide. 


Most overhead of logging mechanisms comes from data copy for creating logs and cache-line flushing by special instructions. Cache-line flushing takes a large portion of the total overhead. Use our evaluation with the above workloads as an example again. On average, the cache-line flushing takes 65\% and 51\% of total overhead for undo logging and redo logging mechanisms respectively. Reducing the overhead of cache-line flushing is the key to enable high performance transaction for persistent memory.

The traditional methods reduce the overhead of cache-line flushing using asynchronous cache-line flushing (e.g., blurring persistent boundary~\cite{7208274} and relaxing persistency ordering~\cite{7208274,mnemosyne_asplos11}). Those methods remove the overhead of cache-line flushing off the critical path, by overlapping cache-line flushing with the transaction. However, the effectiveness of asynchronous cache-line flushing depends on the characteristics of the transaction (e.g., how frequent data updates happen), cache-line flushing can still be exposed into the critical path, increasing the latency of the transaction.

In this paper, we introduce a new method to reduce the overhead of cache-line flushing. The traditional methods work at the program level and leverages program semantics: as long as the transaction semantics remains correct, we can change the order of persisting data and trigger asynchronous cache-line flushing. Different from the traditional methods, our method introduces \textit{architecture awareness}.
In particular, we do not flush certain cache lines, as long as those cache lines are eliminated out of the cache because of the hardware caching mechanism (e.g., the cache replacement algorithm). In other words, we rely on the existing hardware mechanism to automatically and implicitly flush cache lines. The traditional methods do not have architecture awareness. Ignoring the possible effects of the caching mechanism, the traditional methods flush cache lines by explicitly issuing cache flush instructions, even though those cache lines will be soon or have been eliminated out of the cache by hardware. 

Furthermore, we examine the cache line dirtiness to quantify the efficiency of cache-line flushing. The dirtiness of a cache line is defined as the ratio of dirty bytes to total number of bytes in a cache line. Since a cache line is the finest granularity to enforce data persistency, the whole cache line has to be flushed, even though only a few bytes in the cache line are dirty. 
Use our evaluation with the above workloads as an example again:  the average dirtiness of flushed cache lines in Redis and SQLite is 49\% and 49\% for undo and redo logging mechanisms respectively. 
Flushing clean data in a cache line wastes memory bandwidth and decreases the efficiency of cache-line flushing. 

To leverage the architecture awareness to enable high performance transactions, we must address a couple of challenges. First, we must have a software mechanism to reason and decide the existence of cache blocks~\footnote{We distinguish cache line and cache block in the paper. The cache line is a location in the cache, and the cache block refers to the data that goes into a cache line.} in the cache, without hardware modification. The mechanism must be lightweight and allow us to make a quick decision on whether a cache-line flushing is necessary. 

Second, we must provide strong guarantee on crash consistency to implement transactions. Skipping cache-line flushing for some persistent objects raises the risk of losing data consistency for committed transactions. The software mechanism to reason the residence of a cache block in the cache is an approximation to the hardware-based caching mechanism. If the software mechanism skips a cache-line flushing, but the corresponding dirty cache block is still in the cache, then there is a chance that the cache block is inconsistent when a crash happens. We must have a mechanism to detect and correct such inconsistency in persistent memory. 

To address the above two challenges, we introduce \name{} (\textit{Arch}itecture-\textit{a}ware, \textit{p}erformant and \textit{p}ersistent \textit{t}ransaction), an architecture-aware, high performance transaction runtime system. 
\name{} provides a new way to perform transactional updates on  persistent memory with efficient cache-line flushing.
To address the first challenge, \name{} uses an LRU queue to estimate the residence of cache blocks of a persistent object in the cache and decide whether cache flushing  for a persistent object in a transaction is necessary. 

To address the second challenge, \name{} introduces a lightweight checksum mechanism. Checksums are built using multiple cache blocks from one or more persistent objects to establish implicit invariant relationships between cache blocks. Leveraging the invariant, \name{} can detect data inconsistency and make best efforts to correct data inconsistency after a crash happens. The checksum mechanism provides a strong guarantee on crash consistency, while causes small runtime overhead (less than 5\% loss in throughput in our evaluation). 

 
Furthermore, to improve the efficiency of cache-line flushing, we examine the implementation of common database systems  (Redis and SQLite), and find two problems accounting for the low dirtiness of flushed cache lines. The two problems are unaligned cache-line flushing and uncoordinated cache-line flushing. The two problems come from the fundamental limitation of the existing memory allocation mechanism designed for the traditional DRAM. In particular, the existing memory allocation does not consider the effects of cache-line flushing on persistent memory, and spread data structures with different dirtiness across cache blocks. This causes the low dirtiness of flushed cache lines. \name{} introduces a customized memory allocation mechanism to coalesce cache-line flushing and improve efficiency. 


In summary, the paper makes the following contributions:
\begin{itemize}[leftmargin=*]
    \item An architecture-aware new method to achieve high performance transactions on persistent memory; 
    
    \item A mechanism that determines the necessity of cache-line flushing based on the locality of cache blocks; A checksum mechanism to detect and correct data inconsistency to provide strong guarantee on crash consistency;
    
    \item We reveal the low dirtiness of flushed cache lines in two common databases, and provide a solution to improve the efficiency of cache-line flushing;
    
    \item With \name{}, on average we reduce cache-line flushing by 66\% and improve system throughput by 19\% (42\% at most), when running YCSB (A-F) and TPC-C against Redis, and OLTP-bench (TPC-C, LinkBench and YCSB) against SQLite, using a undo logging (PMDK) and a redo logging (Mnemosyne) as baseline. \name{} provides strong crash consistency demonstrated by our crash tests with four hardware caching policies.
\end{itemize}

\vspace{-5pt}
\section{Background and Motivation}
\label{sec:bg}

Many studies build a atomic and durable transaction~\cite{pmdk, atlas:oopsla14, Dulloor:eurosys14, nvheap:asplos11, Kolli:ASPLOS2016, mnemosyne_asplos11, 7208274, cdds:fast11,7208276,Kolli:ASPLOS2016} to handle the crash consistency issue on NVM. With such a transaction, each update must be ``all or nothing'', i.e., either successfully completes, or fails completely with the data in NVM intact. With such a transaction, one has to write back the modified data from the volatile cache to NVM that provides the durability. To ensure a cache line is written to NVM in a correct order, one often uses cache-line flushing instructions (e.g., \texttt{clflush}, \texttt{clflushopt} or \texttt{clwb}) and persistent barriers (e.g., \texttt{sfence} and \texttt{mfence}). Cache-line flushing is expensive, because of two reasons: (1) it may need to invalidate cache lines (with \texttt{clflush} and \texttt{clflushopt} instructions) and trigger cache-line sized writes to the memory controller; and (2) it needs persistent barriers to ensure that all flushes are completed and force any updates in the memory controller to be written to NVM.   



In this paper, we use the term \textit{persistent object} to represent a data object that is modified within the transaction and needs to be persisted. We use the term \textit{log record} to represent a log (an copy of the old data in an undo logging mechanism or a copy of the new data in a redo logging mechanism).  To persist a persistent object, the current common practice is to flush all cache blocks of the persistent object~\cite{pmdk} \footnote{We have to flush all cache blocks, even though some of them may not be in the cache, because there is no mechanism to faithfully track which cache blocks are in the cache.} 
We use ``flushing all cache blocks'' and ``cache-line flushing'' interchangeably to indicate persisting a persistent object.


We do not consider battery-backed RAM as a solution to reduce cache-line flushing. Using battery-backed RAM, it is possible to avoid cache-line flushing to build durable transactions, because the battery allows the system to flush cache lines to persistent memory when a crash happens. However, battery-backed RAM has restrictive temperature requirements, leakage risk, limited storage time, long re-charge cycles, finite battery shelf life, and overall high cost-of-ownership~\cite{batteryNVM:drawback}. 

\subsection{Performance Analysis on Log-based Transactions}
\begin{figure}
\includegraphics[width=0.48\textwidth, height=0.14\textheight] {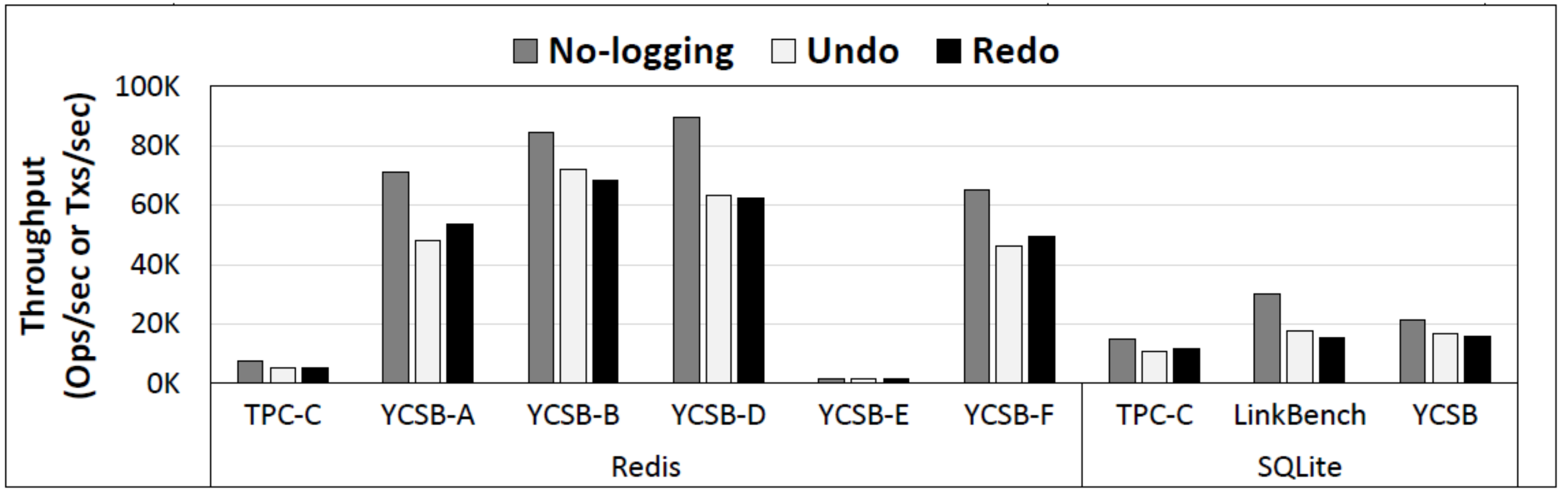}
 \vspace{-5pt}
\caption{Throughput when running YCSB workloads (A-F), TPC-C and LinkBench against Redis and SQLite with and without logging. }
\vspace{-5pt}
\label{fig:undo_redo_overhead}
\end{figure}

Undo and redo logging are two common mechanisms to build persistent transactions on persistent memory. In undo and redo logging, the logging operations (including data copy and log record manipulation) and persistence operations (including cache-line flushing and store barrier) are necessary. Both of them cause performance loss in a transaction. To quantify the impact of the persistent logging on transaction throughput, we run multiple workloads, including YCSB and TPC-C against Redis, and OLTP-bench (TPC-C, LinkBench and YCSB) against SQLite with and without the persistent logging. For each workload, we use eight client threads. Section~\ref{sec:exp_method} has more experiment details. Figure~\ref{fig:undo_redo_overhead} shows the results. 

\begin{figure}[t]
\includegraphics[width=0.48\textwidth, height=0.14\textheight] {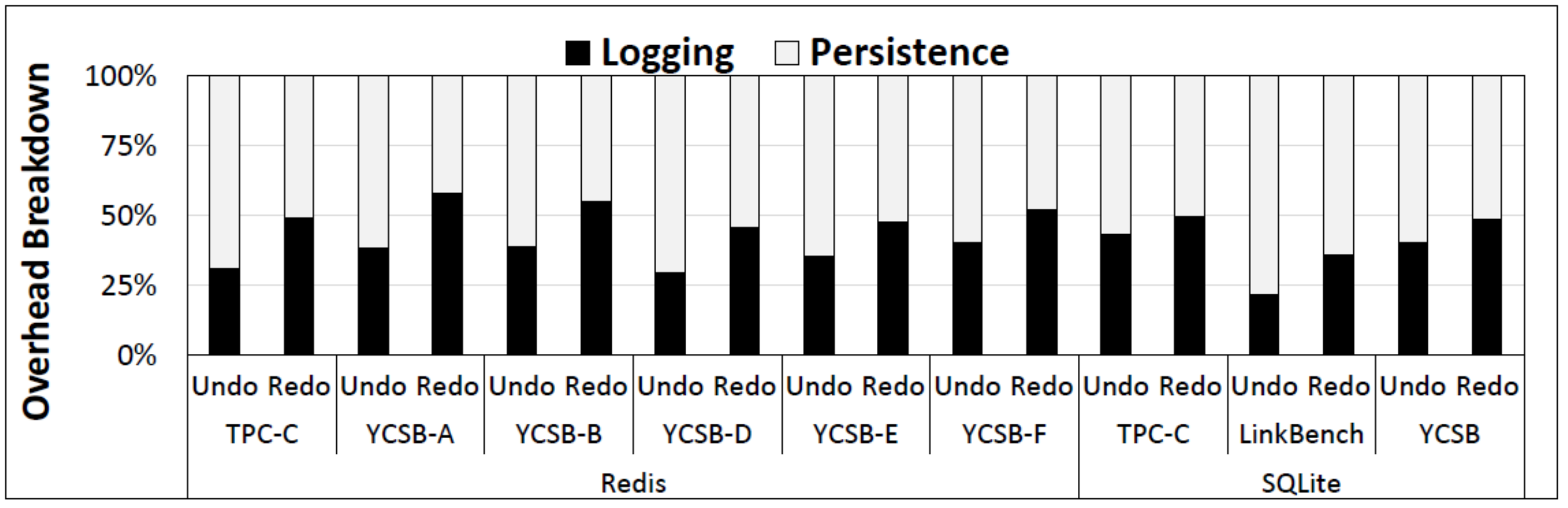}
 \vspace{-5pt}
\caption{Breakdown of undo/redo logging latency overhead. }
\vspace{-15pt}
\label{fig:undo_redo_write_latency_overhead_breakdown}
\end{figure}

The figure reveals that logging decreases throughput by 8\%-49\%. For a workload with frequent updates (YCSB-A) or large updates (LinkBench), the logging overhead can be very large (33\% and 49\% for YCSB-A and LinkBench respectively). Furthermore, we measure the latency overhead caused by logging operations and persistence operations. Figure~\ref{fig:undo_redo_write_latency_overhead_breakdown} shows the results. In the undo and redo logging, the persistence operations account for 56\%-78\% and 41\%-64\% of the latency overhead respectively; 
The overhead of those persistence operations is exposed to the critical path of transactions. The above results show that the persist operations can significantly impact the transaction performance. Thus, we must avoid frequent cache-line flushing.

Introducing architecture awareness into the design of a transaction, we want to skip cache flushing by leveraging data reuse information in the cache. If data reuse is low, then there is a very good chance that the data is eliminated out of the cache by the hardware-based caching mechanism. We study data reuse in the next section. 
\subsection{Data Reuse and Dirtiness Analysis}
Data in a transaction includes log records 
and persistent objects. Log records, which are used to maintain the transaction atomicity, are seldom reused.
We study data reuse at the persistent object level, and explore whether there are persistent objects with few reuse. These persistent objects are candidates for skipping cache-line flushing.

\begin{figure}[t]
\includegraphics[width=0.48\textwidth, height=0.14\textheight] {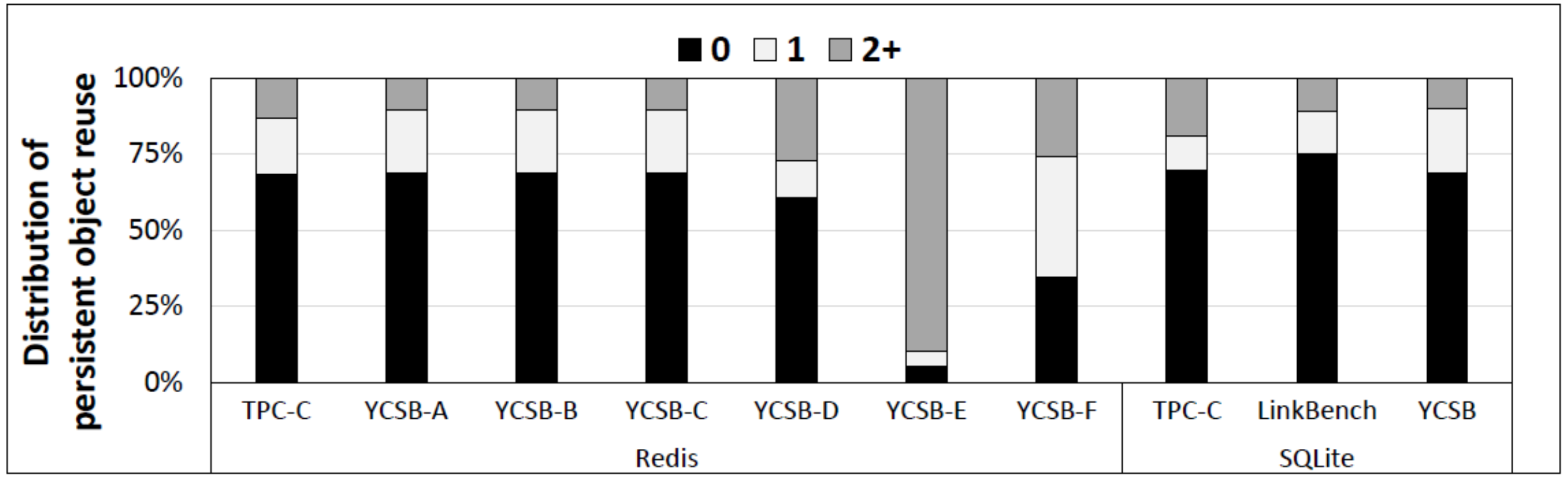}
 \vspace{-5pt}
\caption{Distribution of data reuse for persistent data objects.}
\vspace{-10pt}
\label{fig:ycsb_tpcc_data_reuse}
\end{figure}

To study data reuse, we count the number of operations (read and write) for each persistent object, and then report what is the percentage of persistent objects with 0, 1, 2 or more operations, which we call the distribution of data reuse.
Figure~\ref{fig:ycsb_tpcc_data_reuse} shows the results. 
The figure reveals that 78\% of persistent objects are used only once or twice in all workloads except YCSB-E.
In YCSB-E, about 89\% of persistent objects have data reuse no less than 2. Such high data reuse is because of the following reason: This workload has frequent queries, each of which cover a range of persistent objects. Those ranges of queries overlap with each other, causing high data reuse.

We also explore the efficiency of cache-line flushing. In particular, we quantify the average dirtiness of flushed cache lines. Table~\ref{tab:avg_dirtiness} shows the results for undo and redo logging (the two logging mechanism have the same dirtiness). In general, the dirtiness is less than 0.6 in all workloads, which is low. 

\begin{table}[]
\begin{center}
\scriptsize
 \caption{Average dirtiness of flushed cache lines.}
 \vspace{-5pt}
        \label{tab:avg_dirtiness}
\begin{tabular}{|p{0.6cm}|p{0.3 cm}|p{0.3 cm}|p{0.3 cm}|p{0.3 cm}|p{0.3 cm}|p{0.6cm}|p{1.0cm}|p{0.6cm}|}\hline
\multicolumn{6}{|c|}{\textbf{Redis}}                                                             & \multicolumn{3}{c|}{\textbf{SQLite}}                                                                   \\ \hline
\multirow{2}{*}{\textbf{TPC-C}} & \multicolumn{5}{c|}{\textbf{YCSB}}                             & \multirow{2}{*}{\textbf{TPC-C}} & \multirow{2}{*}{\textbf{LinkBench}} & \multirow{2}{*}{\textbf{YCSB}} \\ \cline{2-6}
                                & \textbf{A} & \textbf{B} & \textbf{D} & \textbf{E} & \textbf{F} &                                 &                                     &                                \\ \hline
0.31                            & 0.55       & 0.55       & 0.51       & 0.51       & 0.56       & 0.40                               & 0.49                                  & 0.46                             \\ \hline

\end{tabular}
\end{center}
\vspace{-20pt}
\end{table}



\textbf{Conclusions.} Using the industry standard workloads, our analysis on data reuse and dirtiness shows great opportunities to enable high performance transactions by skipping cache-line flushing and improving its efficiency. 


\vspace{-5pt}
\section{Design}
\label{sec:design}
Motivated by the above performance analysis, we introduce a high-performance transaction runtime system.

\subsection{Overview}
\name{} avoids cache-line flushing for persistent objects (but not log records) to enable high performance transactions without disturbing transaction atomicity.  \name{} uses an LRU-based method to reason if persistent objects are in the cache. 
With this approach, \name{} does not immediately make a decision on flushing cache blocks for a persistent object, when a cache flushing request is issued from a transaction to persist a persistent object. \name{} delays the decision until it collects more information on read/write of the persistent object and estimates the locality of the persistent object, using the LRU queue. For a persistent object that is estimated to be out of the cache, the cache flushing for all of its cache blocks is skipped.

\name{} is also featured with a checksum mechanism. Skipping cache-line flushing for some persistent objects raises the risk of having inconsistent data for committed transactions, when a crash happens. 
To remove the risk, we introduce a checksum mechanism. This mechanism generates checksums for persistent objects that have cache-line flushing skipped. The checksum mechanism builds invariant relationships between cache blocks. Upon a crash, the checksums are used to detect and correct data inconsistency. We design the mechanism with the consideration of 
avoiding runtime overhead and maximizing the capabilities of correcting data inconsistency. 

Further, we identify two reasons accounting for the low dirtiness of flushed cache lines: unaligned cache-line flushing and uncoordinated cache line flushing. To address the two problems, \name{} introduces a customized memory allocation mechanism. It clusters persistent objects with the same functionality (i.e., key, field, value, or log) into contiguous cache blocks to coordinate and align cache-line flushing, 
based on which \name{} improves the efficiency of cache-line flushing. 

\begin{figure}[!t]
 \vspace{-5pt}
    \centering
    \includegraphics[width=0.5\textwidth, height=0.2\textheight]{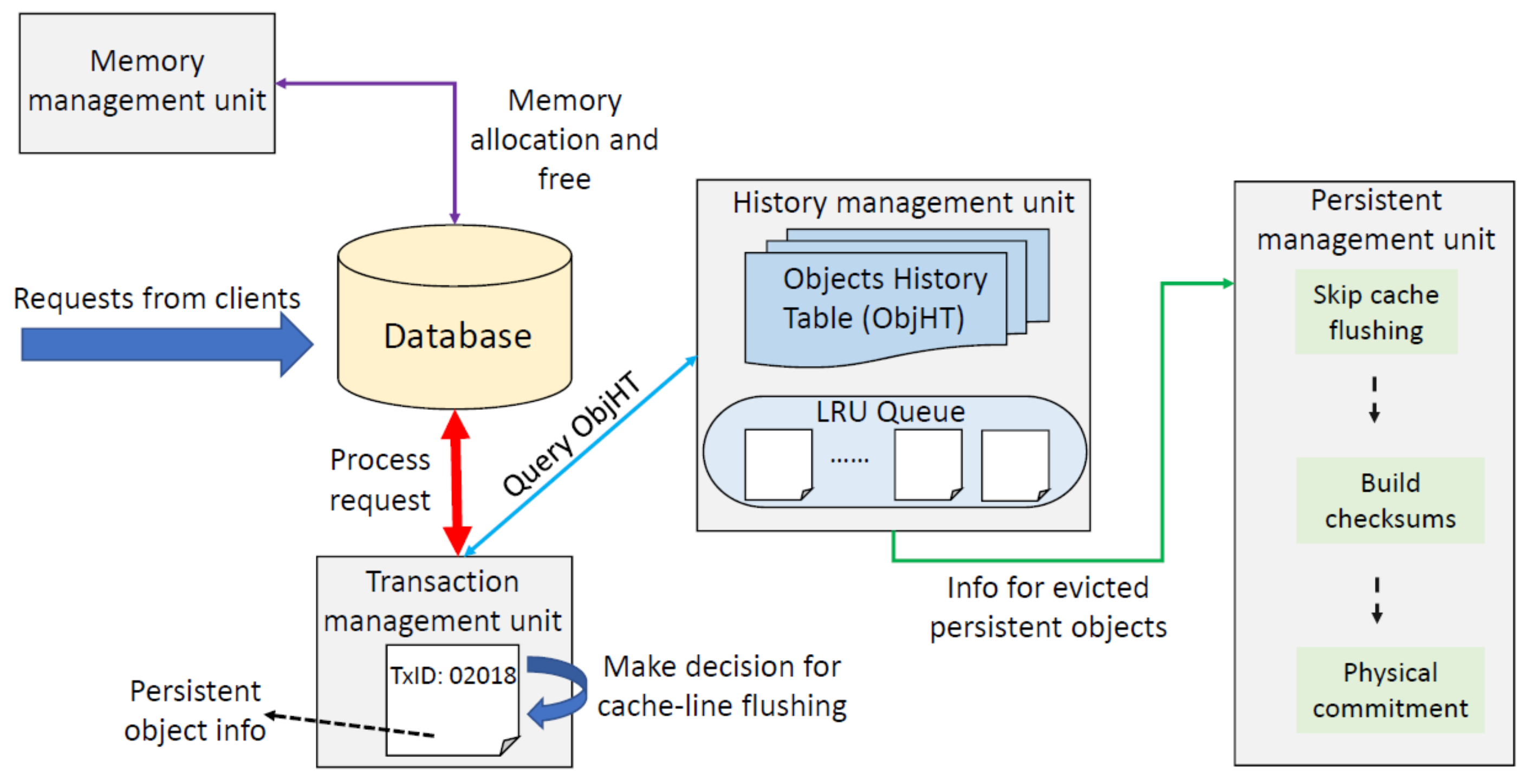}
     \vspace{-10pt}
	    \caption{The architecture of \name{}.} 
    \label{fig:Archa_architecture}
    \vspace{-20pt}
\end{figure}

\textbf{Overall architecture of \name{}.}
\name{} has four major components: transaction management unit, memory management unit,  persistent management unit, and history management unit. Figure~\ref{fig:Archa_architecture} shows the architecture of \name. 

(1) The transaction management unit includes a set of APIs to establish a transaction (i.e., start and end). Such transaction information is sent to the \name{} runtime to implement transaction semantics. The transaction management unit processes the requested operations of the transaction. It also flushes cache blocks for persistent objects that are estimated to be in the cache.
(2) The memory management unit pre-allocates a set of memory pools for coalescing cache blocks and manages the pools to meet memory allocation requests from transactions. 
(3) The persistent management unit builds checksums for persistent objects for which \name{} skips the cache-line flushing. 
(4) The history management unit maintains an LRU queue and a hash table, $ObjHT$. The LRU queue is used to estimate the locality of persistent objects (i.e., in the cache or not). $ObjHT$ is used to provide metadata information for each persistent object in the LRU queue, such as the location in the LRU queue and whether there is any pending cache-line flushing.

\subsection{Architecture-Aware Cache-Line Flushing}
\label{sec:basic_design}
The architecture-aware cache-line flushing uses an LRU
queue to reason if a persistent object is in the cache or not, and skips cache-line flushing for it, if not. When a persistent object is updated, its cache blocks are placed into the LRU queue (the queue length is equal to the capacity of last level cache), and the decision for cache flushing for this persistent object is pending until we have enough information to estimate the residence of the persistent object in the cache,
based on the LRU queue.  We describe our design in details as follows. 

First, once \name{} receives a request (i.e., a read or write operation to a persistent object) from the client, the transaction management unit queries
$ObjHT$ to see if the requested persistent object has a record there. If yes, we infer that the persistent object is accessed recently. The hardware cache may have the persistent object resident in the cache because of a previous operation on the persistent object. If the previous operation is a write operation, flushing cache blocks for the persistent object must have been pending. We finish the cache flushing for the previous write operation. Furthermore, we update the location of the persistent object in the LRU queue, because of the current request. If the current request is a write operation, we hold the cache flushing for the current request, waiting for the opportunity to skip it in the future.


If the transaction management unit cannot find the requested persistent object information in $ObjHT$, we conclude that the persistent object has not been accessed recently. The hardware cache may evict the persistent object out of the cache or never access it at all. The transaction management unit then skips any pending cache flushing request for the persistent object. Afterwards, the transaction management unit asks the history management unit to insert the information for the persistent object into the LRU queue, and suspend the cache flushing for the most recent request if the request writes the persistent object. In the future transactions, as other persistent objects are accessed, the target persistent object can be evicted out of the LRU queue according to the LRU policy, and its record will then be removed from $ObjHT$ and the pending cache flushing will be skipped. 

We must maintain the commit status of a transaction very well. After the completion of a transaction, we cannot label it as commit as in the traditional transaction, because cache flushing for some persistent objects in the transaction may be pending. For such a transaction, we label it as \textit{logical commit}. Only after all of cache flushing for persistent objects in the transaction are either finished or skipped (but with checksums added to the persistent objects. See Section~\ref{sec:checksum}), we label the transaction as \textit{physical commit}.

A logically committed transaction has completed all read and write operations in the transaction. For such a transaction, the system does not respond to the client for transaction commit. 
For a physically committed transaction, the system does so, as in the traditional undo or redo logging mechanisms.

The modern hardware-based cache hierarchy employs sophisticated caching policies. It is possible that a persistent object is resident in the cache while the LRU estimates otherwise. For this case, skipping cache-line flushing can cause data inconsistency for a physically committed transaction, when a crash happens. We introduce a checksum mechanism to detect and correct inconsistent data (see Section~\ref{sec:checksum}). 

In our evaluation (see ``random crash tests'' in Section~\ref{sec:eval_results}), we find that our LRU-based approach tends to be conservative to estimate data locality for our workloads. In particular, evaluating with four hardware caching policies, we find that when a crash happens, there are at most a few tens of inconsistent persistent objects, much smaller than the number of inconsistent persistent objects in the LRU queue. This means that some persistent objects that are estimated to be resident in the cache by the LRU approach are not in the cache. Using such a conservative approach can reduce the number of inconsistent persistent-objects when the crash happens.

\textbf{Handling log records.}
Log records, once created for a transaction, are seldom accessed (unless a crash happens). We could skip cache flushing for log records and rely on the hardware-based caching mechanism to implicitly persist them. However, by doing so, some log records that are not timely flushed by the hardware are lost when a crash happens; We raise the risk of losing transaction atomicity before the physical commitment of the transaction.  
Hence, we do not skip cache-line flushing for log records. They are committed and maintained as in the traditional logging mechanisms. 

\textbf{Memory utilization.} 
Our LRU-based approach requires a queue and a hash table. Maintaining these data structures increases memory footprint (typically a few megabytes, depending on the number of persistent objects estimated to be in the cache). 
This memory overhead can pollute the cache and hence lose performance. However, even with the overhead, \name{} still brings performance benefit, because (1) some operations on the two data structures are not in the critical path, (2) the benefit of skipping cache lines overweigh the overhead, and (3) some workloads have few data reuse and hence are not sensitive to the reduction of the cache capacity.




\subsection{Checksum Design}
\label{sec:checksum}
Skipping cache-line flushing for some persistent objects raises the risk of disturbing transaction atomicity: once a transaction is physically committed, there is no strong guarantee on crash consistency, because we \textit{estimate} data locality and the estimation can be inaccurate. To remove the risk, we introduce a checksum mechanism. 

We have multiple requirements for the checksum design. First, the checksum mechanism should have the capability to detect data inconsistency in physically committed persistent objects. Second, the checksum mechanism must provide strong guarantee on crash consistency for persistent objects when they are physically committed. Third, the checksum mechanism must be lightweight. 
The overhead of the checksum maintenance should be smaller than the performance benefit of skipping cache-line flushing for persistent objects. We describe the design of the checksum mechanism in this section.

\begin{figure*}[t]
  \centering
   \subcaptionbox{Checksum creation}[.32\linewidth][c]{%
    \includegraphics[height=0.16\textheight]{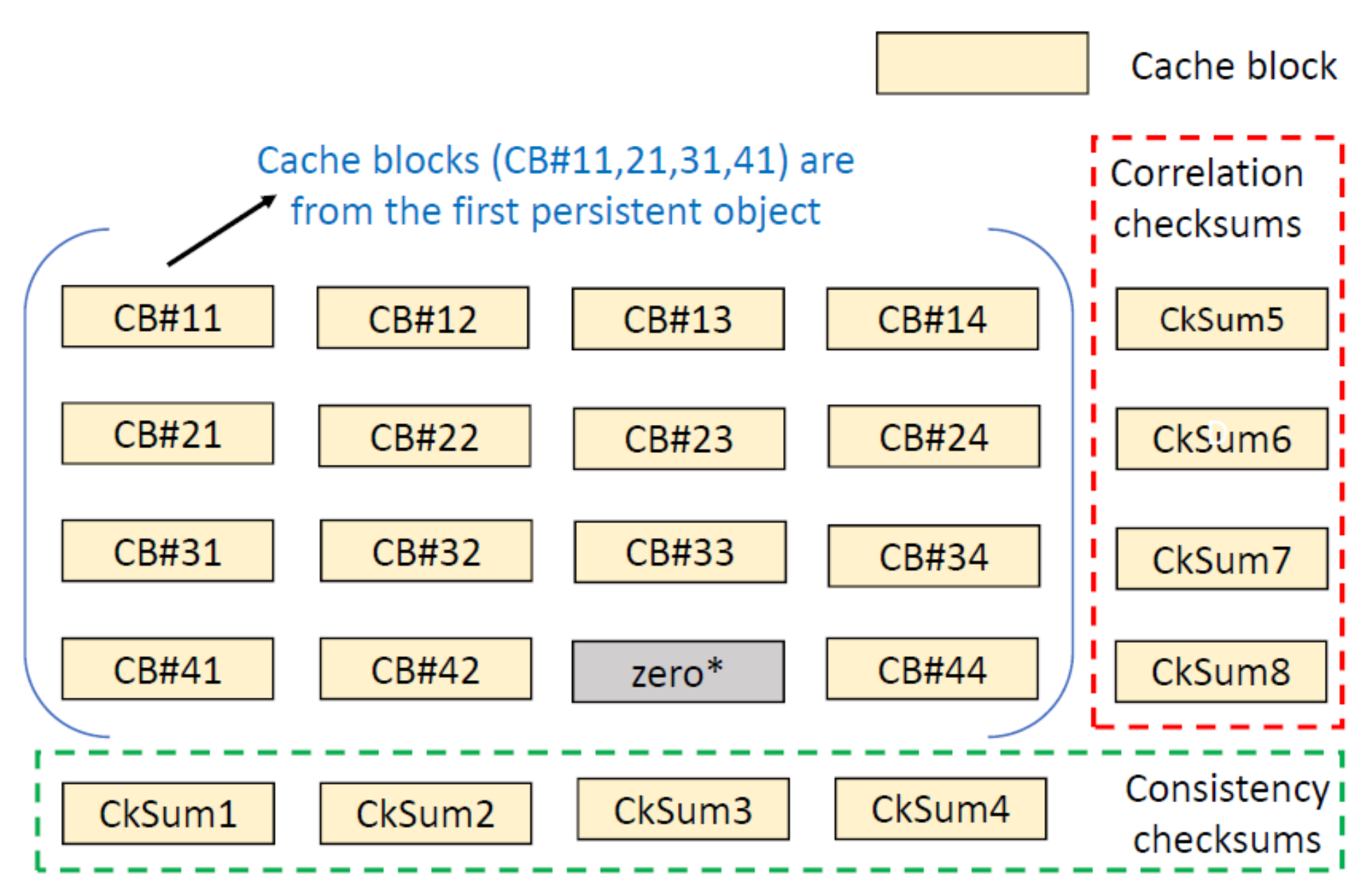}\label{fig:checksum_a}}\hspace{5pt}
  \subcaptionbox{An example of correctable data inconsistency}[.32\linewidth][c]{%
    \includegraphics[height=0.16\textheight]{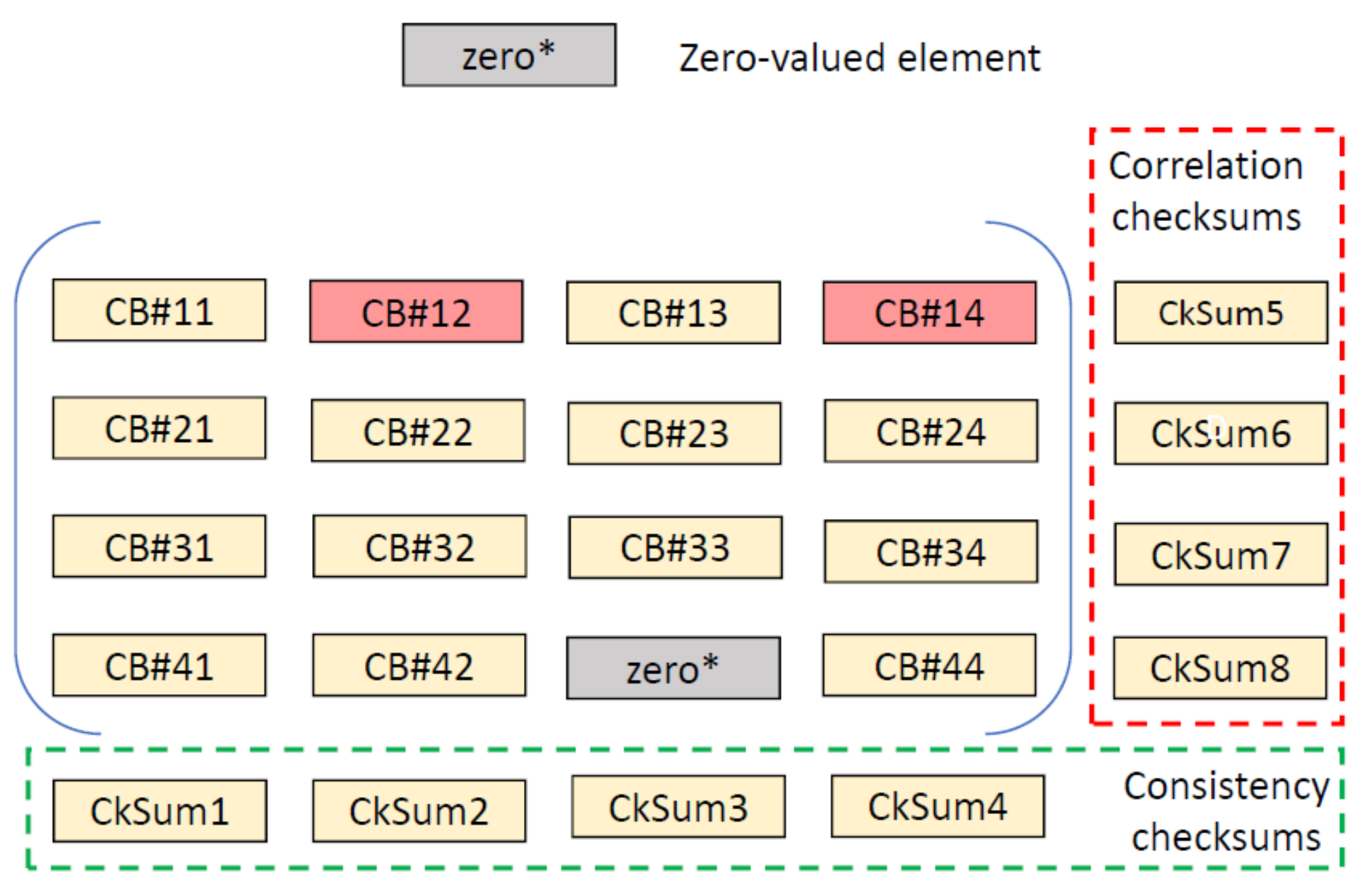}\label{fig:checksum_b}}
    \hspace{5pt}
  \subcaptionbox{An example of uncorrectable data inconsistency}[.32\linewidth][c]{%
    \includegraphics[height=0.16\textheight]{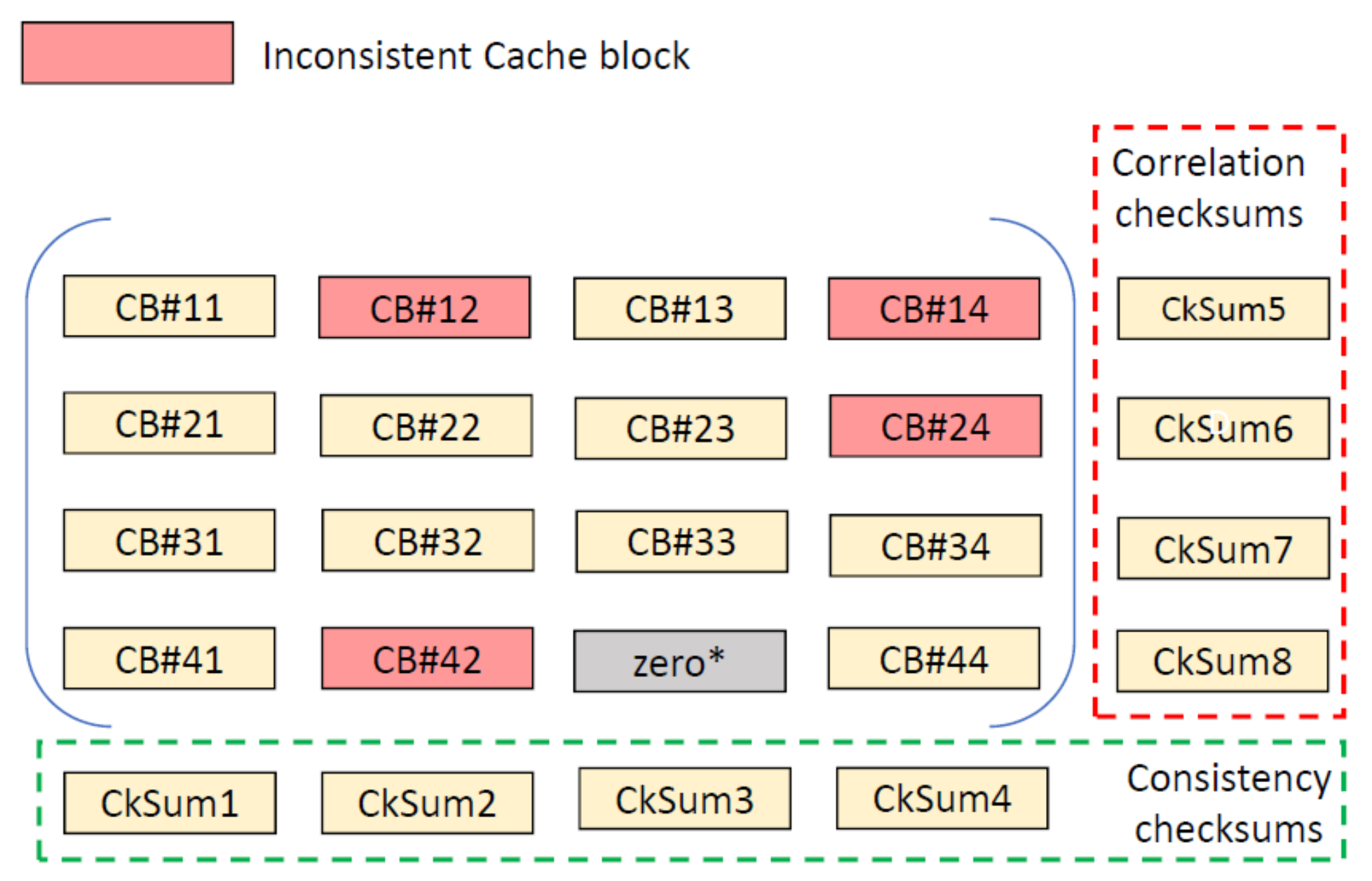}\label{fig:checksum_c}}
    \vspace{-5pt}
  \caption{Three examples for checksum creation and correcting data inconsistency.}
   \vspace{-15pt}
    \label{fig:checksum}
\end{figure*}

\textbf{General Description.}
Our checksums are built with cache blocks of multiple persistent objects from one or more transactions. 
To build the checksums, cache blocks of multiple persistent objects are logically organized as an $M \times N$ matrix ($M$ and $N$ are the dimension sizes of the matrix, discussed later). Those persistent objects have cache-line flushing skipped. Each column of the matrix corresponds to cache blocks of one or more persistent objects, where each element of the column is a cache block. Checksums are built as one extra row (the $(M+1)_{th}$ row)  and one extra column (the $(N+1)_{th}$ column) of the matrix. The matrix becomes $(M+1) \times (N+1)$. 
The extra row, named \textit{consistency checksums}, is used to detect and correct data inconsistency of $N$ columns, 
and each element of the extra row is a consistent checksum for one column. The extra column, named \textit{correlation checksums}, builds an invariant relationship between cache blocks across multiple persistent objects. 
The correlation checksums can correct data inconsistency. We name the matrix,  \textit{virtual matrix}. 
 

\textbf{Consistency checksums to detect data inconsistency.}
When a persistent object with cache-line flushing skipped is logically committed, we immediately create a consistency checksum. 
The checksum is a simple summation of cache blocks of the persistent object. 
The checksum is immediately flushed for persistency once it is created. 
When a crash happens, for each persistent object with a consistency checksum, we recalculate the checksum and compare it with the existing one in persistent memory. If there is a mismatch, then data inconsistency is detected. 



The consistency checksum mechanism is very effective to detect data inconsistency for a persistent object. Any cache block of the persistent object with data inconsistency can easily cause checksum mismatch. In our evaluation running ten workloads consisting of billions of 
transactions and 40,000 crash tests under four hardware caching polices, the consistency checksum mechanism detects all inconsistent data. 


\textbf{Correlation checksums to correct data inconsistency.}
A correlation checksum, as an element of the $(N+1)_{th}$ column of the virtual matrix, is a summation of cache blocks of a row in the virtual matrix. The $(N+1)_{th}$ column is composed of $M$ correlation checksums, each of which is for one row. Since the cache blocks of a row come from at most $N$ persistent objects, the correlation checksum at the end of the row aims to correct data inconsistency for any persistent object in the row. Correlation checksums (i.e., the $(N+1)_{th}$ column) are immediately flushed out of the cache for persistency, once they are fully built.

Once a crash happens, we recalculate correlation checksums and compare them with the existing ones in persistent memory. If there is a mismatch in any correlation checksum (e.g., the element $m_{k(N+1)}$ of the virtual matrix), then the corresponding row (the row $k$ in the example) must have data inconsistency. Using consistency checksums, we can reason which element of the row $k$ is inconsistent. Assume that the element $m_{kj}$ is. This element is corrected by the following:
\begin{equation}
\scriptsize
\label{eq:correlation_checksum}
m_{kj} = m_{k(N+1)} - \sum_{i=1, i \neq j}^{N} m_{ki}
\end{equation}
\noindent where $m_{k(N+1)}$ is the correlation checksum committed in persistent memory.

We pack persistent objects into the virtual matrix following the order they are allocated. A persistent object can use more than one columns of the virtual matrix, if one column is not big enough to hold all cache blocks of the persistent object. A column, after populated by cache blocks of a persistent object, can have zero-valued elements, if the persistent object is not big enough and the next persistent object is too big to be placed into the column. Persistent objects are allocated by \name{}, from memory pools pre-allocated before transactions happen (Section~\ref{sec:coalesce_cachelines}). This method allows \name{} to control over the assignment of virtual addresses to persistent objects to implement the virtual matrix-based checksum mechanism.

\textbf{An example.}
Figure~\ref{fig:checksum}.a shows an example to further explain the idea of checksums. In this example, we have four persistent objects with four, four, three, and four cache blocks respectively. The virtual matrix is $4 \times 4$, and each column has one persistent object. The consistency checksums are in the fifth row, and the correlation checksums are in the fifth columns. The consistency checksums, $CkSum1$-$ChkSum4$ can detect data inconsistency for the first-fourth persistent objects respectively. The correlation checksums, $CkSum5$-$CkSum8$, can be used to correct data inconsistency for the cache blocks in the first-fourth rows. 
Suppose $CB\#32$ has inconsistency detected by the consistency checksum $CkSum2$. The inconsistency can be corrected by the correlation checksum $CkSum7$. In particular, $CB\#32 = CkSum7 - CB\#31 - CB\#33 - CB\#34$.


\textbf{Locating checksums.}
When a persistent object is updated, we must locate its related checksums to update them. We introduce a lightweight approach to correlate persistent objects with their checksums, such that we can quickly locate checksums for any given persistent object.


Consistency and correlation checksums are placed into the virtual memory page where the persistent objects protected by the checksums reside, such that using the virtual addresses of the persistent objects, we can easily locate the checksums.

In particular, we fix the virtual matrix size as $8\times8$ ($N=M=8$). A memory page with the memory page size of 4KB (annotated as $mps$) has only one virtual matrix. The first 3136 bytes of the memory page (i.e., $7\times7$ cache blocks) is allocated for cache blocks of persistent objects, which is annotated as $oms$. 
The remaining 896 bytes (two $7\times1$ cache blocks) of the memory page are for the consistency and correlation checksums, which is annotated as $cms$. 
Given a virtual memory address of a persistent object ($obj\_addr$) and its size, we use the following two equations to locate its checksum. The equations calculate the addresses of consistency and correlation checksums ($cons$ and $corr$) respectively.

\vspace{-5pt}
\footnotesize
\begin{equation}
\label{eq:consistency_chksum}
cons = (\ceil{(obj\_offset \: / \: (cms \: / \: 2)} - 1)  \times cbs + oms 
+ psa
\end{equation}
\vspace{-20pt}


\begin{equation}
\label{eq:correlation_chksum}
corr = 
(obj\_offset \: \% \: (cms \: / \: 2) - cbs + oms + (cms \: / \: 2) \\
+ psa
\end{equation}
\normalsize
where $obj\_offset$ is the distance between the starting address of the persistent object and the starting address of the memory page in which the object reside ($obj\_offset$ = $obj\_addr$ \% $mps$), $cbs$ is the cache block size, and $psa$ is the starting address of the memory page ($psa$ = $obj\_addr$ - $obj\_offset$).


In essence, by putting checksums at a fixed page offset, we simplify the efforts to locate checksums. 
The above algorithm to locate checksums is based on the assumption that a persistent object is never larger than a memory page, which is true in all workloads we evaluate. If the persistent object is larger than a memory page, then we build a larger virtual matrix and use multiple memory pages (instead of one) to hold it, and put checksums at a fixed offset of the multiple memory pages.

\textbf{High performance checksum.}
Our checksum mechanism does not cause large performance overhead because of the following reason. Creating checksums is not in the critical path of a transaction. A checksum for a persistent object is created, only after the persistent object is estimated to be evicted out of the LRU queue. This indicates that it is highly likely that the persistent object will not be accessed in the near future. Hence it is highly likely that creating checksums does not block operations on the persistent object. Also, creating checksums for the persistent object and committing the checksums do not block the execution of other transactions. Hence, checksum creation can happen in parallel with other operations, which removes it from the critical path.


Second, checksums do not need to be updated frequently. Once a persistent object is updated, its checksums must be recalculated and updated to maintain the validness of checksums. Such updates can cause performance overhead. This performance problem is common in other mechanisms, such as ECC or RAID. However, it is not a problem in our design, because we use those persistent objects that are not frequently accessed (according to the LRU queue) to build checksums. Updating checksums does not happen often. 


Third, the overhead of flushing cache blocks of checksums can be smaller than that of flushing cache blocks of persistent objects in the traditional undo/redo logging-based methods. This fact is supported by our performance evaluation (Section~\ref{sec:eval_results}).
Given an $N \times M$ virtual matrix, we need to flush ($N + M$) cache blocks to make checksums consistent. In contrast, to make persistent objects in the virtual matrix consistent in the traditional methods, we flush at least ($N$) cache blocks (this case happens when each column has one small data object with the size of one cache block), and at most ($M \times N$) cache blocks (this case happens when each column has a large data object with the size of $M$ cache blocks).

If the virtual matrix is populated well with cache blocks of large data objects, then the number of cache blocks to flush for checksums is significantly smaller than that for persisting persistent objects (i.e., $N + M$ vs. $M \times N$). Hence, we reduce the number of cache-line flushing, even though we need to flush checksums. However, if the virtual matrix is not populated well and each column has only a small data object, then the number of cache blocks to flush for checksums can be larger than that for persisting $N$ persistent objects (i.e., $N+M$ vs. $N$). To handle such cases with small data objects, we pack multiple small objects into the same column to heavily populate the virtual matrix.

\textbf{Analysis on the capability of correcting data inconsistency.}
The correlation checksums have a strong capability to correct data inconsistency. If a cache block in a row is inconsistent, we can easily correct it using Equation~\ref{eq:correlation_checksum}. If more than one cache block in a row are inconsistent, we can use the consistency checksums \textit{and} the correlation checksums to correct them. The consistency checksum is not only used to detect data inconsistency, but also used to correct the inconsistency of cache blocks that fall into the same column, using the similar method as in Equation~\ref{eq:correlation_checksum}.

Figure~\ref{fig:checksum}.b gives an example where we have two inconsistent cache blocks in the first row. Using the correlation checksum $CkSum5$ alone is not able to correct them.
However, using the consistency checksums $CkSum2$ or $CkSum4$, we can correct at least one inconsistent data. Afterwards, we can use $CkSum5$ to correct the other. 


It is possible that a row has multiple inconsistent cache blocks and the columns where those inconsistent cache blocks reside have another inconsistent cache blocks.  Figure~\ref{fig:checksum}.c gives an example. In this example, the first row have two inconsistent cache blocks ($CB\#12$ and $CB\#14$). They reside in the columns two and four. These two cache blocks are not correctable by the correlation checksum $CkSum5$. Meanwhile, each of the columns two and four has another inconsistent cache block ($CB\#42$ and $CB\#24$), making the consistency checksums ($CkSum2$ and $CkSum4$) incapable of correcting the inconsistent cache blocks too.

In this case, any checksum, including the combination of consistency checksum and correlation checksum, cannot correct those cache blocks. However, such a case is extremely rare: Those inconsistent cache blocks must be so ``coincident'' to fall into the same row and column together. In our evaluation with ten workloads including 
billions of transactions and 400,000 crash tests under four hardware caching polices, our checksums can correct all of data inconsistency for committed transactions.

\textbf{Post-crash processing.}
After a crash happens, we examine persistent objects in persistent memory. If they do not have checksums and the transactions are physically committed, then the persistent objects must be consistent without any cache flushing skipped. 
If they do not have checksum and the transactions are not physically committed, then the transaction updates are cancelled and the persistent objects are restored using traditional logs. 

If the persistent objects have checksums and the transactions are physically committed, then we use consistency checksums to detect consistency of each persistent object. If there is any inconsistency, then we use correlation and consistency checksums to correct them.  If the data inconsistency is not correctable, 
which is very rare, then the corresponding transaction update is lost.
To avoid incorrectable data inconsistency after physical commitment, we could add another row and column as consistency and correlation checksums. The new and old checksums, each of which is built upon half of rows or columns, can enhance correction capabilities for those rare cases, but come with larger performance overhead. Study of this tradeoff is out of scope of this paper, because the current checksum mechanism already works well in our evaluation. 

\textbf{Ensuring transaction atomicity.} Before the physical commitment, \name{} relies on logs, as in the traditional undo and redo logging mechanisms, to ensure atomicity. After the physical commitment, the persistent objects are successfully updated with the assists of checksums and the atomicity is enforced. In the extremely rare case where the persistent object is not consistent and the checksum mechanism cannot correct it, we rely on a traditional checkpoint mechanism and go back to the last valid checkpoint to ensure atomicity.

\definecolor{codegreen}{rgb}{0,0.6,0}
\definecolor{codegray}{rgb}{0.5,0.5,0.5}
\definecolor{codepurple}{rgb}{0.58,0,0.82}
\definecolor{backcolour}{rgb}{0.95,0.95,0.92}
\lstdefinestyle{style1}{
    commentstyle=\color{codegreen},
    keywordstyle=\color{magenta},
    numberstyle=\tiny\color{codegray},
    stringstyle=\color{codepurple},
    basicstyle=\footnotesize,
    frame=single,
    numbers=left,                    
    numbersep=5pt, 
	escapeinside={(*@}{@*)},
}
\lstset{style=style1}

\subsection{Coalescing of Cache-Line Flushing}
\label{sec:coalesce_cachelines}

To reduce the overhead of cache-line flushing, we have two methodologies: one is to avoid cache flushing for persistent objects as in Section~\ref{sec:basic_design}; the other is to coalesce cache-line flushing to avoid low dirtiness in flushed cache lines. 
After investigating two common databases (Redis and SQLite), 
we find two reasons accounting for low dirtiness: 
unaligned cache-line flushing and uncoordinated cache-line flushing.

The unaligned cache-line flushing happens when a persistent object is not aligned with cache lines. For example, a persistent object is 100 bytes. Ideally, the object should use two cache blocks (assuming that the cache block size is 64 bytes). However, the object could not be aligned well during the memory allocation, and uses three cache blocks. Once the object is modified, we have to flush three cache lines instead of two. This easily increases the number of cache-line flushing by 50\%. We find this problem in Redis and SQLite.


The uncoordinated cache-line flushing happens when multiple, associated data objects are allocated into separate cache blocks.
The multiple data objects are associated, because they are often updated together. If they are allocated into the same cache blocks, then we can reduce the number of cache-line flushing. This problem happens more often in NoSQL systems, such as a key-value store system. We use Redis as an example to explain it further. 

As a key-value store system, Redis enables secondary indexing based on a two-level hash table. In the second level, Redis has a set of field-value pairs.  
For each field-value pair, the field and value objects are allocated separately on different cache blocks. This is inefficient on persistent memory, because the field and value have to be persisted by flushing separate cache lines. The size of the field object is small (usually less than one cache block), and the field object is usually updated with the value object together. Therefore, coalescing the field and value objects into a fewer contiguous cache blocks can reduce the number of cache-line flushing.
To address the above problems, we introduce a memory allocation mechanism to improve the efficiency of cache-line flushing. 
The original implementation of Redis uses the traditional memory allocation,  without considering the implications of memory allocation on cache flushing. Whenever a key or a value is created, Redis allocates the corresponding memory space on demand, without the coordination with other memory allocations. 
In our new design for Redis, we pre-allocate three memory pools without allocating memory on demand. The three memory pools,  $key\_pool$, $field\_value\_pool$ and $log\_pool$, meet the memory allocation requests for keys, field and value pairs, and log records, respectively. We use the three memory pools, instead of allocating memory on demand, because the three pools can use separate memory allocation methods to minimize cache-line flushing; The three memory pools can also cluster objects with the same functionality (i.e., key, field, value, or log) into contiguous cache blocks to coordinate cache-line flushing.

\vspace{-5pt}
\section{Implementation}

\textbf{Programming APIs.}
\name{} is implemented as a user-level library to provide persistence support and be integrated with the existing log-based transaction implementation, such as Intel PMDK~\cite{pmdk} and Mnemosyne~\cite{Mnemosyne:ASPLOS11}. 
 \name{} includes a set of APIs, defined in Table~\ref{tab:interface}. 

$\name{}\_init()$ is used to pre-allocate multiple memory pools for coalescing cache-line flushing (Section~\ref{sec:coalesce_cachelines}) and initialize critical data structures (e.g., the LRU queue and $QbjHT$). $\name{}\_Tx\_Start()$ and $\name{}\_Tx\_End()$ are used to identify transactions for the \name{} runtime, and can be embedded into the existing transaction start/finalization functions. $\name{}\_Tx\_LCommit()$ is used to replace the traditional transaction commit to implement the logical commit for \name{}. $\name{}\_Malloc()$ and $\name{}\_Free()$ are used to replace the traditional memory allocation and free APIs in the transaction implementation. The two APIs are used to allocate and free memory from/to the pre-allocated memory pools for coalescing cache-line flushing.


\begin{table}
        \begin{center}
       \scriptsize
               \caption{\name{} APIs}
                \vspace{-5pt}
                 \label{tab:interface}
        \begin{tabular}{|p{4.2cm}|p{3.5cm}|}
        \hline
        \textbf{API Name}     & \textbf{Functionality}                                 \\ \hline \hline
        {\fontfamily{qcr}\selectfont \name{}\_Init()}       & Pre-allocate memory pools and initialization         \\\hline
        {\fontfamily{qcr}\selectfont \name{}\_Tx\_Start()}      & Identify the beginning of a transaction             \\ \hline
        {\fontfamily{qcr}\selectfont \name{}\_Tx\_End()}       & Identify the end of a transaction                 \\ \hline
        {\fontfamily{qcr}\selectfont \name{}\_Tx\_LCommit()}       &    Logical commitment            \\ \hline
        {\fontfamily{qcr}\selectfont Archa\_Malloc(int type, size\_t size)}    & Memory allocation for coalescing cache-line flushing                    \\ \hline
        {\fontfamily{qcr}\selectfont \name{}\_Free(int type, size\_t size)}   & Free memory allocation for coalescing cache-line flushing    \\ \hline
        \end{tabular}
        \end{center}
        \vspace{-21pt}
\end{table}

\textbf{System optimization.}
\name{} includes a number of optimization techniques to enable high performance and thread safety. These techniques include SIMD vectorization of checksum creation and update, a high-performance concurrent lock-free hash table, and a high-performance LRU queue based on circular buffers. In addition, to avoid contention on the LRU queue from multiple transactions (multiple threads), \name{} creates a transaction management unit for each transaction, and the transaction management unit puts information on write/read of persistent objects into a local buffer. By fetching the information from those local buffers, the history management unit collectively updates the LRU queue.

\vspace{-5pt}
\section{Evaluation}
\vspace{-5pt}

\subsection{Experimental Methodology}
\label{sec:exp_method}



The goal of the evaluation is to evaluate the performance of \name{} with a range of workloads with different characteristics. We use both NoSQL and SQL systems (Redis and SQLite). We use seven persistent transaction mechanisms for evaluation: undo logging and redo logging with \name{}, undo logging and redo logging without \name{}, the rollback journal system in SQLite, the AOF mechanism (logging every write operation) in Redis, and Atlas~\cite{Atlas:HP} an undo-logging-based system that provides durability guarantees for failure-atomic sections). 
Our implementation of the undo logging and redo logging are based on Intel PMDK~\cite{pmdk} and Mnemosyne~\cite{Mnemosyne:ASPLOS11} (using \textit{asynchronous cache-line flushing}) respectively. For the rollback journal and AOF, whenever a transaction commitment happens, we commit the transaction updates to memory (not hard drive), in order to enable fair performance comparison. 



We run YCSB~\cite{YCSB} (A-F) and TPC-C~\cite{TPCC,pytpcc} against Redis, and run OLTP-bench~\cite{oltpbench} (particularly, TPC-C, LinkBench~\cite{linkbench:simod13} and YCSB) against SQLite. These workloads are chosen for Redis and SQLite respectively, because they can easily run on the two database systems without any modification. Table~\ref{tab:bench_info} gives some details for these workloads. For YCSB running against Redis, we perform transaction operations on 10M key-value pairs, and each of the pair size is 1KB. We choose default data access distributions (uniform or zipfian); For TPC-C aginist Redis, we use 10 warehouses and run five minutes per test. For other workloads against SQLite, we use default configurations. For those experiments using Redis, we run both servers and clients on a single machine as in~\cite{optane:ucsd,ido:micro18}. 

\begin{table}[]
\scriptsize
        \begin{center}
            \caption{The percentage of different operations in evaluated workloads; ``R'', ``U'', ``I'', ``RU'', ``S'' and ``D'' standard for read, update, insert, read \& update, scan, and delete operations respectively.}
        \label{tab:bench_info}
  \vspace{-5pt}
\begin{tabular}{|p{0.2 cm}|p{0.5cm}|p{0.2 cm}|p{0.2 cm}|p{0.2 cm}|p{0.2 cm}|p{0.2 cm}|p{0.2 cm}|p{0.5cm}|p{0.7cm}|p{0.5cm}|}
       \hline
\multirow{3}{*}{\textbf{}} & \multicolumn{7}{c|}{\textbf{Redis}}                   & \multicolumn{3}{c|}{\textbf{SQLite}}                                        \\ \cline{2-11} 
                         Ops     & \multirow{2}{*}{TPCC} & \multicolumn{6}{c|}{YCSB}    & \multirow{2}{*}{TPCC} & \multirow{2}{*}{LinkBH} & \multirow{2}{*}{YCSB} \\ \cline{3-8}
                              &                        & A  & B  & C   & D  & E  & F  &                        &                            &                       \\ \hline
R                             & 8                      & 50 & 95 & 100 & 95 & -  & 50 & 8                      & 64                         & 50                    \\ \hline
U                             & 47                     & 50 & 5  & -   & -  & -  & -  & 47                     & 16                         & 10                    \\ \hline
I                             & 45                     & -  & -  & -   & 5  & 5  & -  & 45                     & 12                         & 5                     \\ \hline
RU                            & -                      & -  & -  & -   & -  & -  & 50 & -                      & -                          & 10                    \\ \hline
S                             & -                      & -  & -  & -   & -  & 95 & -  & -                      & 4                          & 15                    \\ \hline
D                             & -                      & -  & -  & -   & -  & -  & -  & -                      & 4                          & 10                    \\ \hline
\end{tabular}
\end{center}
\vspace{-22pt}
\end{table}

All experiments are performed on a 24-core machine with two 12-core Xeon Gold 6126 processors with 187GB memory and 19.25MB last level cache. We use DRAM to emulate NVM, since NVDIMM is not on the market at the time of preparing this manuscript. For other slower NVMs, the benefits of \name{} would only be larger because of the reduction of cache-line flushing. 
We use \texttt{clwb} and \texttt{clflushopt} instructions to flush cache lines. These two instructions are the most recent instructions especially designed for high performance cache-line flushing. In our study, the transaction performance of using the two instructions is pretty close to each other (less than 5\% performance difference, much smaller than the performance benefit offered by Archapt), hence we only present the results of using \texttt{clwb} for cache-line flushing.

\vspace{-5pt}
\subsection{Experimental Results}
\label{sec:eval_results}
\textbf{Micro-benchmark Evaluation.}
We use a micro-benchmark to compare the overhead of flushing cache blocks of persistent objects (without checksums as in the traditional undo/redo logging) and creating/updating checksums. We aim to quantify and study the overhead of our checksum mechanism. 

\begin{figure}
\includegraphics[width=0.48\textwidth, height=0.14\textheight] {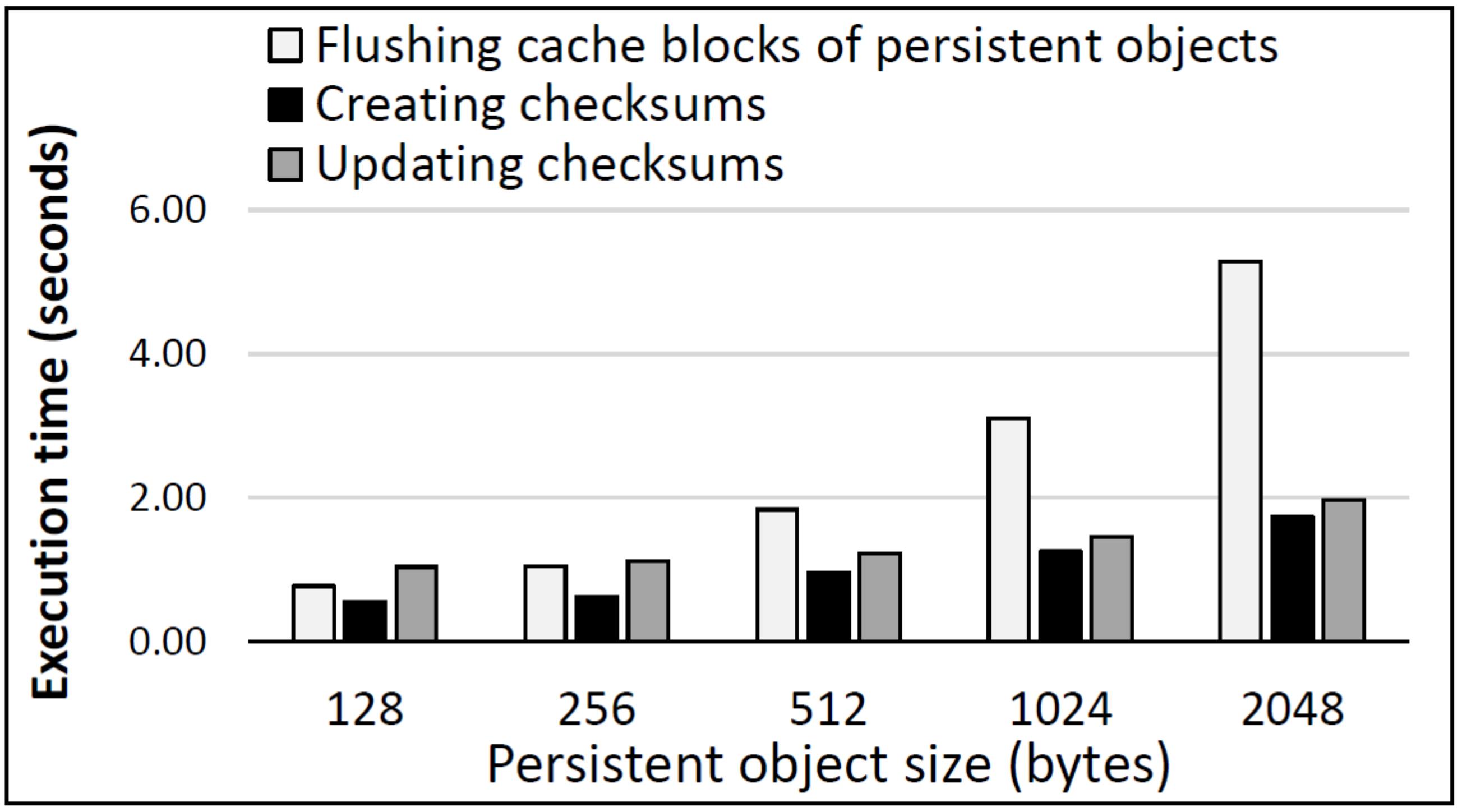}
 \vspace{-5pt}
\caption{Comparing the overhead of flushing cache blocks of persistent objects without checksum and creating/updating checksums.}
\vspace{-15pt}
\label{fig:micro-bench}
\end{figure}

In the micro-benchmark, we randomly access 1.3M persistent objects. We perform five groups of tests. In each group, all persistent objects have the same size, but five groups use different data object sizes. In each group, we perform three tests: (1) Flushing cache blocks of all persistent objects, as in the traditional methods; (2) creating checksums for each persistent object; and (3) updating checksums for each persistent object. Figure~\ref{fig:micro-bench} shows execution time for each test.


Figure~\ref{fig:micro-bench} reveals that the overhead of creating checksums is smaller than that of flushing cache blocks in the traditional methods. 
Creating checksums is cheaper by 29\% and 68\% for small data objects (128 bytes) and large data objects (2048 bytes) respectively. For small persistent objects, packing them into the same column effectively reduces the checksum creation overhead.

Figure~\ref{fig:micro-bench} reveals that for large data objects (2048 bytes), the overhead of updating checksums is smaller (by 63\%) than that of flushing cache blocks in the traditional methods. For small objects (128 bytes), updating checksums is slightly more expensive. However, updating checksums does not happen frequently for persistent objects with checksums (Otherwise, those objects should be estimated to be in the cache and do not have checksums).

\begin{figure*}[!t]
 \centering
 \begin{subfigure}[b]{1.0\linewidth}
    \includegraphics[width=1.0\textwidth, height=0.14\textheight]{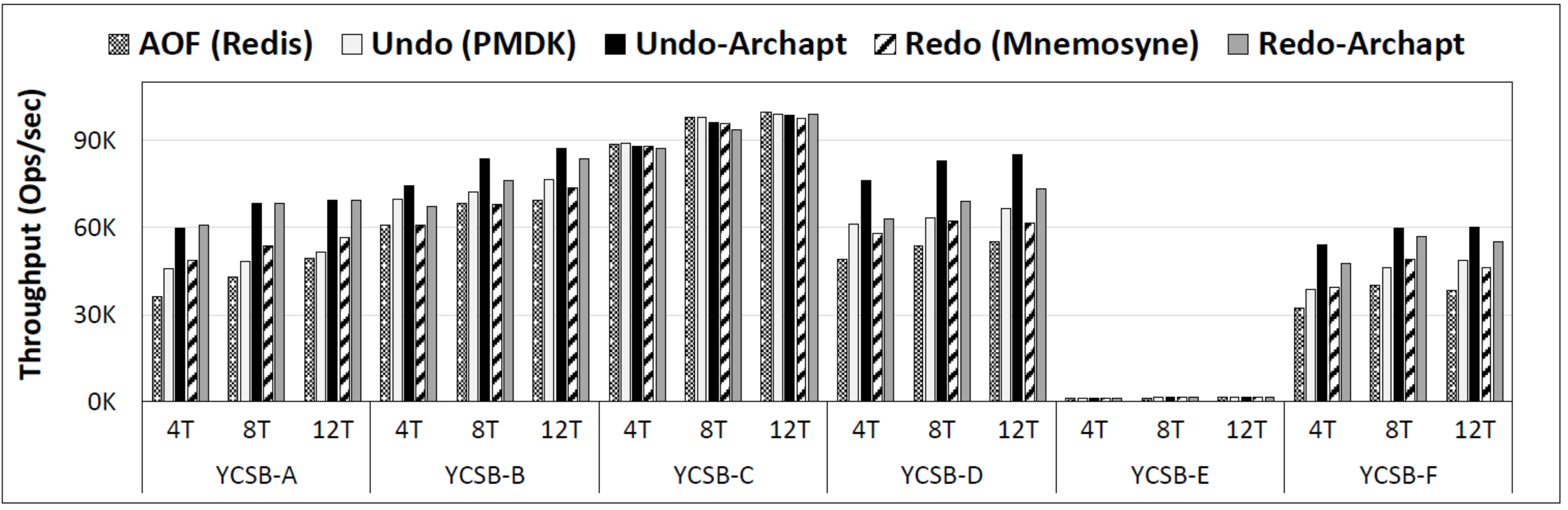}
\label{fig:throughput_redis_ycsb}
\vspace{-15pt}
 \subcaption[a]{YCSB A-F (Redis).}
\end{subfigure}
\begin{subfigure}[t]{0.32\textwidth}
    \includegraphics[width=\textwidth, height=0.14\textheight]{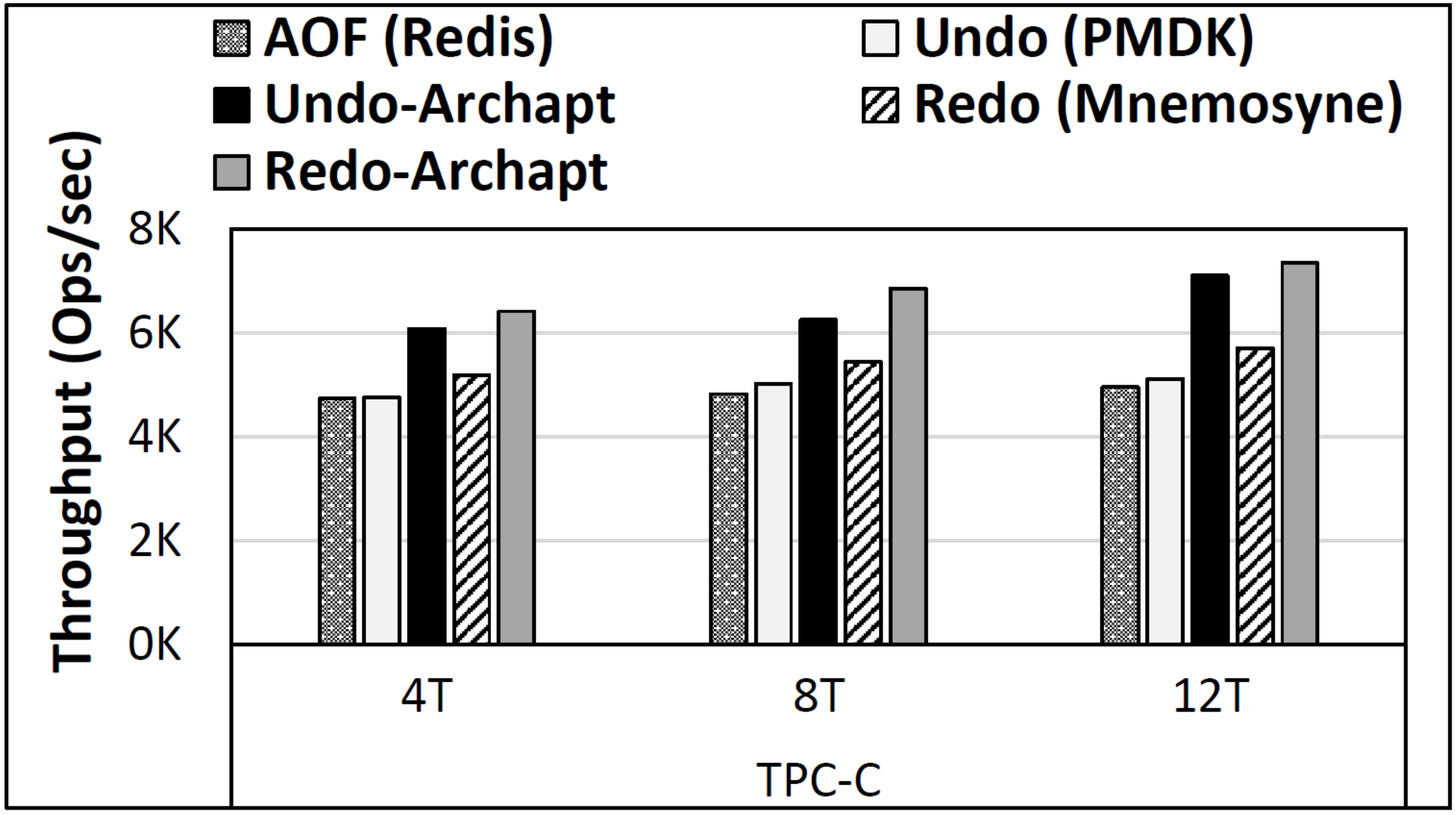}
    \vspace{-15pt}
    \caption{TPC-C (Redis).}
    \label{fig:throughput_redis_tpcc}
  \end{subfigure}\hfill
  \begin{subfigure}[t]{0.635\textwidth}
    \includegraphics[width=\textwidth, height=0.14\textheight]{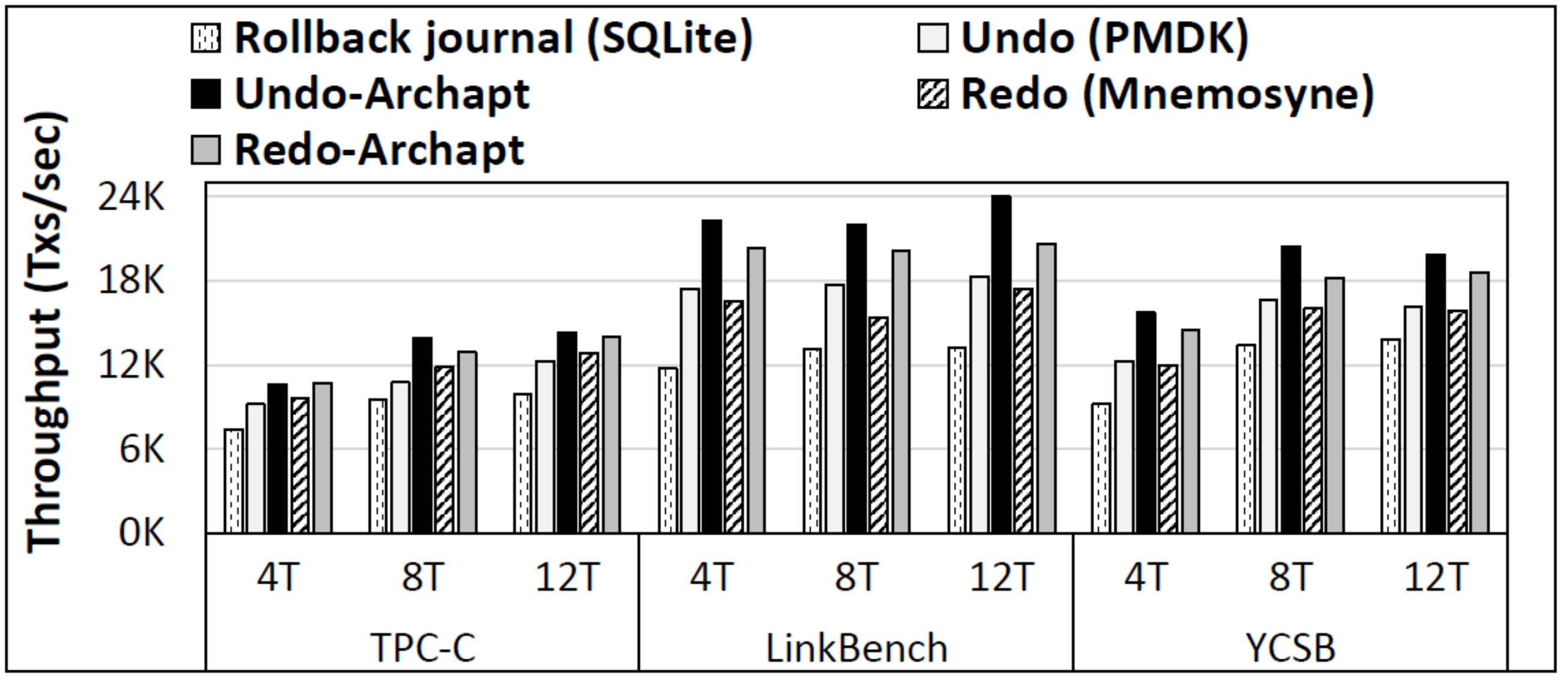}
    \vspace{-15pt}
    \caption{Three workloads with SQLite.}
    \label{fig:throughput_sqlite}
  \end{subfigure}
  \vspace{-5pt}
 \caption{Throughput with the four transaction mechanisms, as the number of threads vary from four to twelve. ``T''=``threads''.}
\label{fig:throughput}
\end{figure*}

\begin{figure*}
\centering
  \begin{subfigure}[b]{0.58\textwidth}
    \includegraphics[width=\textwidth, height=0.14\textheight]{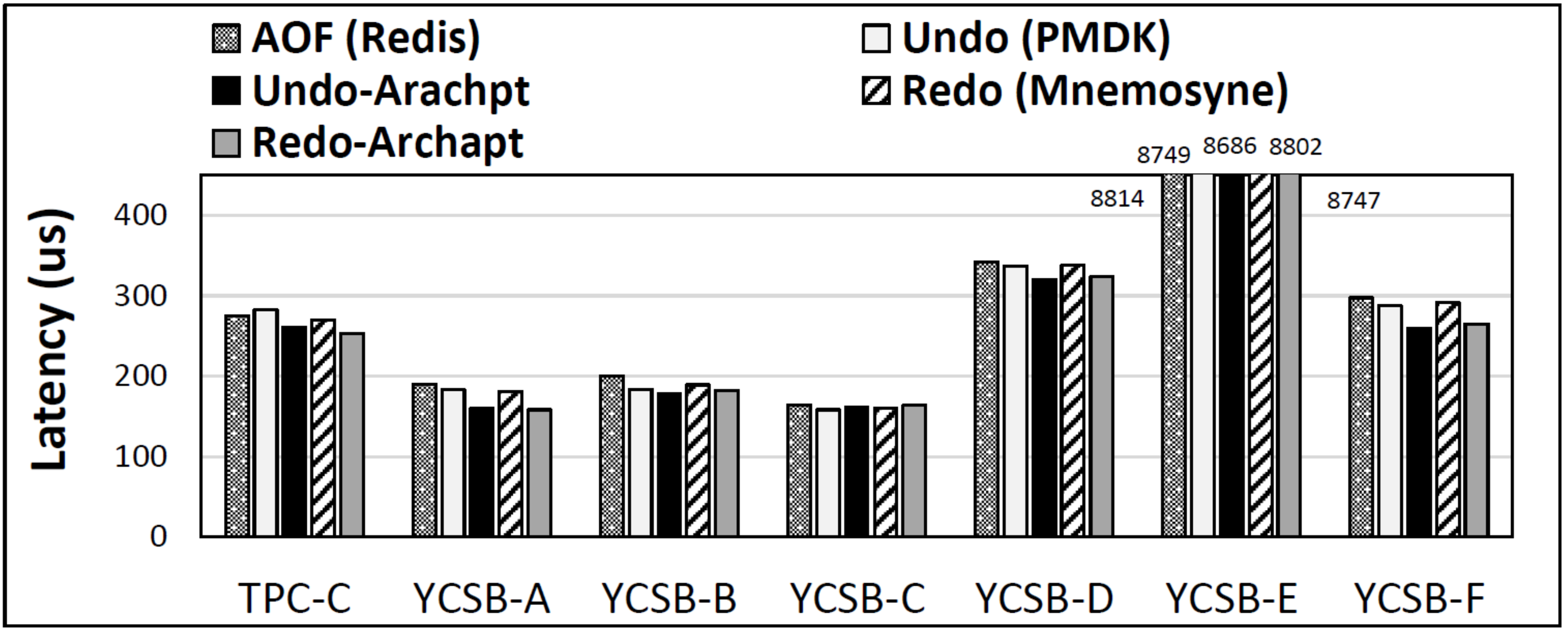}
      \vspace{-15pt}
    \caption{Redis}
    \label{latency_redis}
  \end{subfigure}
  \hfill
  \begin{subfigure}[b]{0.38\textwidth}
    \includegraphics[width=\textwidth, height=0.14\textheight]{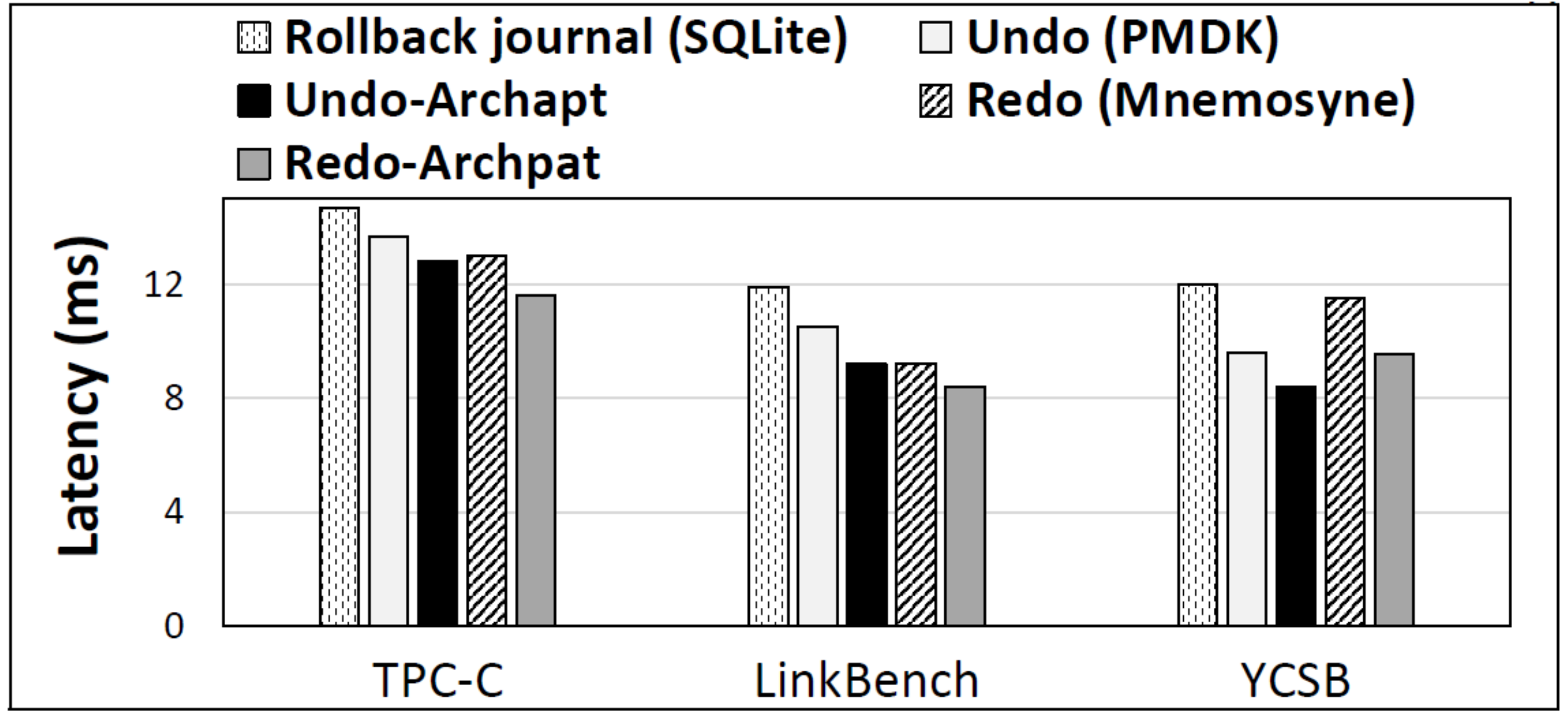}
      \vspace{-15pt}
    \caption{SQLite}
    \label{latency_sqlite}
  \end{subfigure}
  \vspace{-5pt}
  \caption{99th-percentile transaction latency, as the number of threads is eight.} 
  \vspace{-10pt}
  \label{fig:lantency}
\end{figure*}

\textbf{Basic Performance.}
We use different numbers of threads to evaluate throughput and latency. Figures~\ref{fig:throughput} and~\ref{fig:lantency} show the results. We do not include the results for Atlas because of the space limitation, but we discuss the Atlas results as follows.

Figure~\ref{fig:throughput} reveals that the systems with \name{} have higher throughput than all other systems. On average, the system with \name{} offers 54\% and 32\% higher throughput than the rollback journal system in SQLite and AOF in Redis, respectively; For the redo logging (Mnemosyne) and undo logging (PMDK), the system with \name{} increases throughput by 16\% and 22\%, respectively.  


The biggest improvement happens on YCSB-A and LinkBench with undo-\name{}. 
YCSB-A is write-intensive (see Table~\ref{tab:bench_info}) and LinkBench has large persistent objects, which have higher demands on consistency and provide better opportunities for performance improvement. In the best cases, the undo logging with \name{} offers up to 32\%, 54\% and 67\% higher throughput than the other three mechanisms (undo (PMDK), AOF (Redis) and Rollback journal (SQLite)). For the read-only workload (YCSB-C), \name{} cannot offer any performance benefit, because no cache flushing is needed in the original workload, but \name{} provides at most 3\% lower throughput than other mechanisms, which is small.

We also evaluate the performance of Atlas. 
Atlas uses undo logging. Comparing with PMDK using undo logging and \name{}, Atlas performs 9\% and 31\% worse. Atlas has bad performance, because of its mechanism to detect dependencies across failure-atomic sections and instrumentation on persistent operations.

Figure~\ref{fig:lantency} shows the 99th-percentile latency.  We run eight client threads for each workload. On average, \name{} decreases the tail latency by 8\%, 9\%, 9\% and 22\%, compared with the traditional undo logging, redo logging, AOF and the rollback journal system respectively. Such performance improvement in the tail latency comes from the reduction of unnecessary cache flushing. For the read-only workload (YCSB-C) that offers no opportunity to reduce cache flushing, \name{} still provides comparable performance to the other transaction mechanisms. As for Atlas (not shown in the Figure), its transaction tail latency is 5\% and 13\% longer than PMDK  and \name{} respectively.

\begin{figure}
\vspace{-1pt}
\includegraphics[width=0.48\textwidth, height=0.14\textheight] {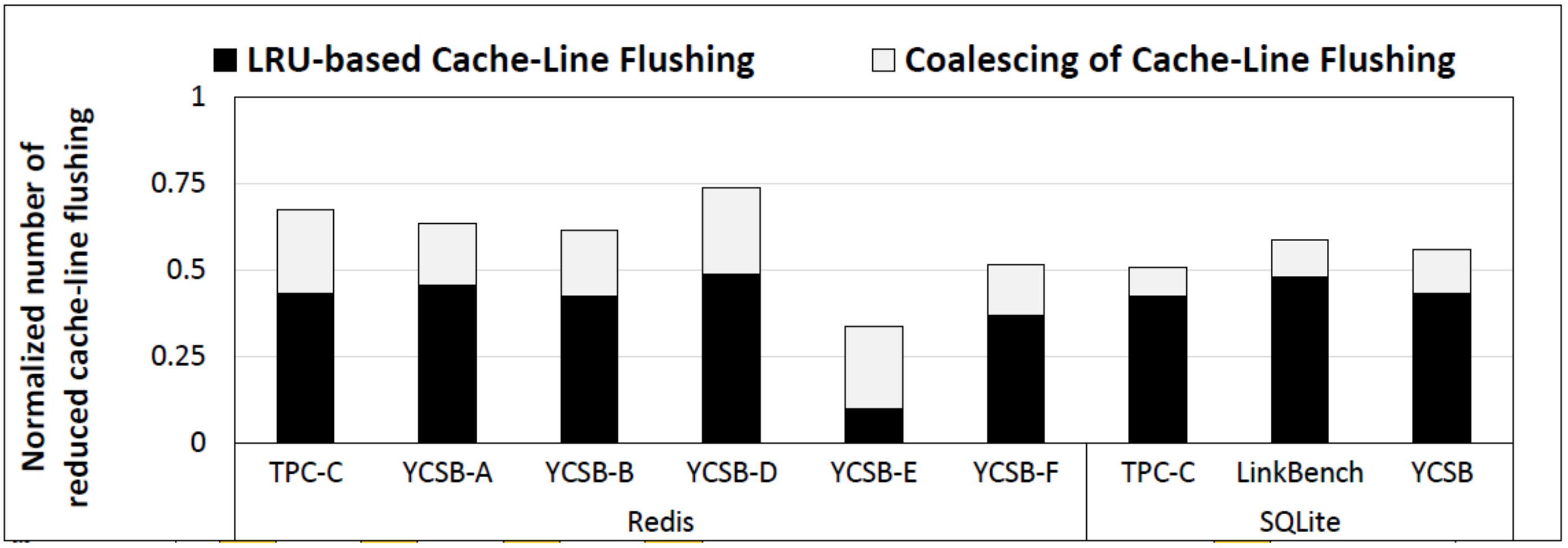}
 \vspace{-5pt}
\caption{The numbers of reduced cache-line flushing after applying \name{} to undo logging. The numbers are normalized by the total numbers of cache-line flushing before using \name{}.}
\vspace{-15pt}
\label{fig:num_cache_line}
\end{figure}

\textbf{Quantifying the effectiveness of reducing cache-line flushing.}
We measure the number of cache-line flushing before and after applying \name{} to undo logging. 
We only show the results for undo logging, because it has less cache-line flushing than the rollback journal and AOF. Hence reducing cache-line flushing for undo-logging is more challenging than doing that for the rollback journal and AOF. Also, we do not show the results for redo logging, because undo and redo logging have the similar number of cache-line flushing.  Figure~\ref{fig:num_cache_line} shows the number of reduced cache-line flushing after applying \name{}. The numbers in the figure are normalized by the total numbers of cache-line flushing before applying \name{}. The figure does not include YCSB-C, because this workload is read-only and does not need cache flushing. The figure also isolates the contributions of the two techniques (the LRU-based approach and coalesce of cache-line flushing) to compare the effectiveness of the two techniques. 

The figure reveals that \name{} greatly reduces the number of cache-line flushing by 66\% on average. YCSB-E has less reduction in the number of cache-line flushing than other workloads, because it has more data reuse in persistent objects, which provides less opportunities to skip cache-line flushing. This result is consistent with that shown in Figure~\ref{fig:ycsb_tpcc_data_reuse}. 

We further notice that both techniques effectively reduce cache-line flushing. The contribution of the LRU-based approach to the reduction of cache-line flushing varies between different workloads, because different workloads have different data reuse of persistent objects. Furthermore, comparing with Redis, SQLite gains less benefit from the coalesce of cache-line flushing. This is because of the strict SQL data structures in SQLite that have some existing optimization on cache line alignment.


\begin{figure}
\includegraphics[width=0.48\textwidth, height=0.14\textheight] {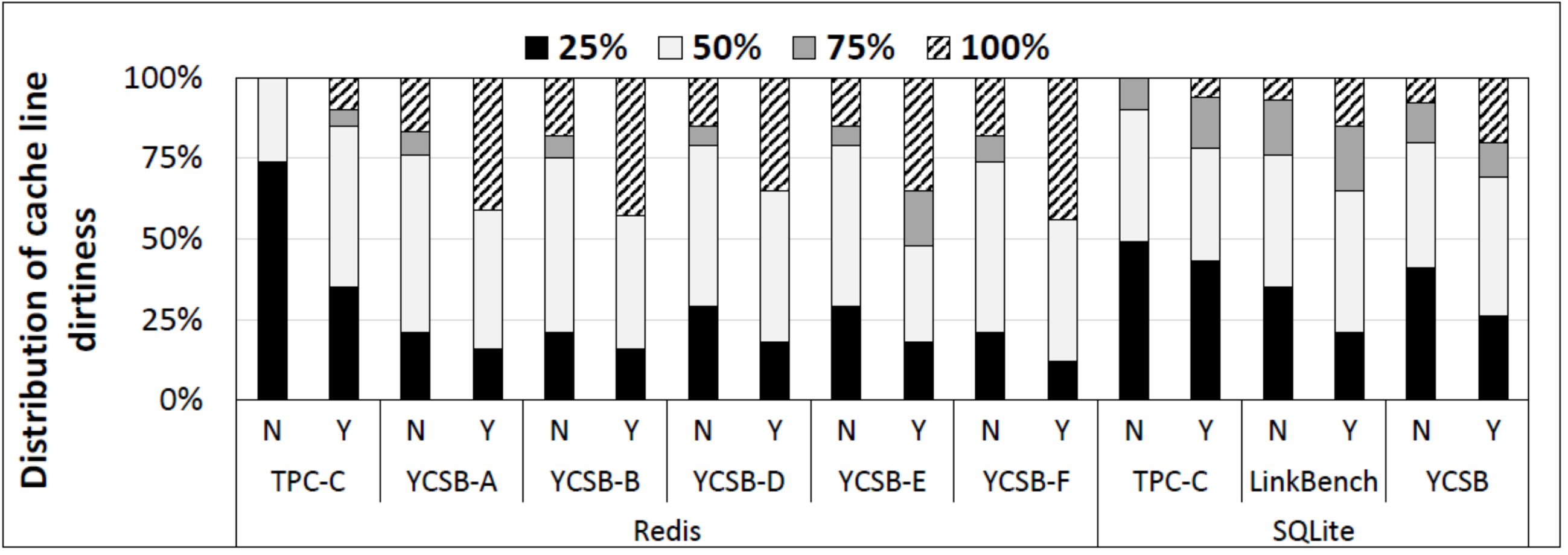}
 \vspace{-5pt}
\caption{Distribution of the dirtiness of flushed cache lines with and without \name{}; ``Y'' and ``N'' standard for using \name{} and no \name{} respectively.}
\vspace{-15pt}
\label{fig:cache_dirtiness_opt}
\end{figure}

\textbf{Quantify dirtiness of flushed cache lines.}
Figure~\ref{fig:cache_dirtiness_opt} shows the distribution of the dirtiness of flushed cache lines before and after applying \name{}. The figure does not include YCSB-C, because this workload is read-only and does not need cache flushing. In general, our memory allocation improves dirtiness by 12\% on average. Among all workloads, YCSB-E (Redis) has the largest increase: the average cache line dirtiness increases from 51\% to 68\%. 


\textbf{Random crash tests. }
\begin{table*}[ht]
\scriptsize
\caption{Crash test results. The numbers for each workload in the table are \textbf{for 100 crash tests}; ``I-obj'', ``DI-obj'' and ``CC-obj'' standard for number of inconsistent persistent-objects, number of inconsistent persistent-objects detected by checksums, and number of inconsistent persistent-objects that cannot be corrected by checksums respectively.}
\label{tab:data_loss_4pl}
\begin{tabular}{|c|c|c|c|c|c|c|c|c|c|c|c|c|c|}

\hline
\multirow{2}{*}{\textbf{Systems}} & \multirow{2}{*}{\textbf{Workloads}} & \multicolumn{3}{c|}{\textbf{Caching policy: LRU}}                                                                                                                                                                                     & \multicolumn{3}{c|}{\textbf{Caching policy: Pseudo-LRU}}                                                                                                                                                                                   & \multicolumn{3}{c|}{\textbf{Caching policy: Bimodal insertion}}                                                                                                                                                                                     & \multicolumn{3}{c|}{\textbf{Caching policy: random}}                                                                                                                                                                                  \\ \cline{3-14} 
                                  &                                     & \textbf{I-obj} & \textbf{DI-obj} & \textbf{CC-obj} & \textbf{I-obj} & \textbf{DI-obj} & \textbf{CC-obj} & \textbf{I-obj} & \textbf{DI-obj} & \textbf{CC-obj}&
                        \textbf{I-obj} & \textbf{DI-obj} & \textbf{CC-obj}  \\ \hline
\multirow{7}{*}{\textbf{Redis}}   & \textbf{TPC-C}                      & 539                                            & 539                                                                  & 0                                                                                            &      504                                          &  504                                                                     & 0                                                                                             &           429                                     &                       429                                               & 0                                                                                             &     798                                          &     798                                                                & 0                                                                                             \\ \cline{2-14} 
                                  & \textbf{YCSB-A}                     &                                  901           &                   901                                                & 0                                                                                             &      776                                          &                   776                                                   & 0                                                                                             &     88                                           &  88                                                                     & 0                                                                                             &  1435                                              &   1435                                                                   & 0                                                                                             \\ \cline{2-14} 
                                  & \textbf{YCSB-B}                     & 0                                              & 0                                                                    & 0                                                                                             &  0                                              &               0                                                       & 0                                                                                             &    0                                            &                 0                                                     & 0                                                                                             &    0                                            &     0                                                                 & 0                                                                                             \\ \cline{2-14} 
                                  & \textbf{YCSB-D}                     & 0                                              & 0                                                                    & 0                                                                                             &                                   0             &               0                                                       & 0                                                                                             &      0                                          &                   0                                                   & 0                                                                                             &  0                                              &   0                                                                   & 0                                                                                             \\ \cline{2-14} 
                                  & \textbf{YCSB-E}                     & 0                                           & 0                                                                 & 0                                                                                             &     0                                           &              0                                                        & 0                                                                                             &     0                                           &                  0                                                    & 0                                                                                            &    0                                            &    0                                                                  & 0                                                                                             \\ \cline{2-14} 
                                  & \textbf{YCSB-F}                     & 29                                            & 29                                                                  & 0                                                                                             &     24                                           &                                               24                       & 0                                                                                             &       7                                        &                  7                                                    & 0                                                                                             &     56                                           &    56                                                                  & 0                                                                                             \\ \hline
\multirow{3}{*}{\textbf{SQLite}}  & \textbf{TPC-C}                      & 782                                            & 782                                                                  & 0                                                                                             &   713                                             &   713                                                                   & 0                                                                                             &  335                                               &   335                                                                   & 0                                                                                             &          941                                      &    941                                                                  & 0                                                                                             \\ \cline{2-14} 
                                  & \textbf{LinkBench}                  & 605                                            & 605                                                                  & 0                                                                                             &       583                                         &                    583                                                  & 0                                                                                             &  305                                               &              305                                                        & 0                                                                                             &    889                                            &                   889                                                   & 0                                                                                             \\ \cline{2-14} 
                                  & \textbf{YCSB}                       & 45                                            & 45                                                                  & 0                                                                                             &      38                                          &                38                                                      & 0                                                                                             &    25                                            &              25                                                        & 0                                                                                             &   96                                            &                96                                                      & 0                                                                                             \\ \hline
\end{tabular}
\vspace{-15pt}
\end{table*}
We examine data consistency in \textit{physically committed} transactions using random crash tests. We aim to evaluate the effectiveness of our checksum mechanism. We use an NVM crash emulator~\cite{Ren:2018:UAR:3286475.3286476},  because of two reasons: (1) a large number of crash tests affect the reliability of our physical machine (\textit{including those battery-backed machines}); and (2) DRAM (used for our NVM emulation) and caches, lose data when the crash happens. The emulator is based on PIN~\cite{pintool} and emulates a configurable LRU cache hierarchy. 
The crash emulator retains data in the emulated main memory after a crash is triggered, allowing us to examine data consistency. The crash emulator randomly triggers crashes. In our evaluation, the cache capacities (19.25 MB in the last level cache) and associativity (11) in the crash emulator are the same as those in our physical machine. For each workload we perform crash tests 100 times to ensure statistical significance.
The emulator only supports the LRU caching policy. We add three more common caching policies (pseudo-LRU, bimodal insertion and random eviction).

Table~\ref{tab:data_loss_4pl} shows the results. The table reports total number of inconsistent persistent-objects measured by the crash emulator and total numbers of inconsistent persistent-objects detected and corrected by the checksum mechanism for 100 crash tests. We have two observations. First, the checksum mechanism successfully detects and corrects \textit{all} inconsistent persistent-objects for all workloads under \textit{the four caching polices}. This demonstrates that the checksum mechanism is highly effectiveness. Second, we do not have a large number of inconsistent persistent-objects after crashes: Given billions of persistent objects to update, we have at most hundreds of inconsistent persistent-objects in each of 100 crash tests. For YCSB-B, YCSB-D, and YCSB-E, we do not even find any inconsistent persistent-objects. This indicates that our LRU-based approach successfully estimates data locality, such that skipping cache-line flushing causes a small number of inconsistent persistent objects after transactions are physically committed.

\vspace{-10pt}
\section{Related Work}
  \vspace{-5pt}

Persistency in NVM has received significant research activities recently~\cite{pmdk, atlas:oopsla14, Dulloor:eurosys14, nvheap:asplos11, Kolli:ASPLOS2016, mnemosyne_asplos11, 7208274, cdds:fast11,7208276,Kolli:ASPLOS2016,Atomic_ucsd:eurosys17} 
Some of them employ undo logging~\cite{pmdk, atlas:oopsla14, Dulloor:eurosys14, nvheap:asplos11, Kolli:ASPLOS2016,7208276}, while others employ redo logging~\cite{mnemosyne_asplos11, 7208274, cdds:fast11}.  Our work can be applied to them to improve transaction performance. 

\textbf{Enabling crash consistency on NVM.} 
Strict persistency~\cite{Pelley:2014:MP:2665671.2665712} enforces crash consistency by strictly enforcing write orders in persistent memory
and can cause a large performance loss.
Some work~\cite{Pelley:2014:MP:2665671.2665712,micro15:joshi,kolli:micro16,Condit:sosp09,7856636}  
relaxes the constrains on write orders to improve performance. The recent work~\cite{buffer_cache_nvm:idpps17} proposes a software cache solution to improve performance by buffering writes to NVM.
Different from the above existing work, we do not relax write orders, but optimize performance by skipping and coalescing cache-line flushing and provide strong guarantee on data consistency for crashes.

\textbf{Detection and correction of data errors. }
Previous efforts on RAID~\cite{Chen:1994:RHR:176979.176981,868660,Menon:1993:AFC:165123.165144,Schmuck:2002:GSF:1083323.1083349} and ECC~\cite{8567428,abft_ecc:SC13,lotecc_isca12,mage_sc12,Yoon:2010:VFE:1736020.1736064,chipkill_tech_rp} exploit hardware- and software-based approaches to detect and correct data errors, but the relatively large runtime overhead and hardware modifications make them hard to be applied to detect and correct data inconsistency on NVM for a transaction mechanism. 


Algorithm-based fault tolerance, as an efficient software mechanism to correct data errors, has been used for fault tolerance in high performance computing~\cite{8048960, 21-chen2011algorithm, jcs13:wu, ics11:davies, hpdc13:davies, ppopp13:chen, 1676475, Du:2012:AFT:2145816.2145845}. However, they are customized for specific numerical algorithms, and are hard to be applied to transactional workloads in database.

\textbf{Probabilistic crash consistency.} 
In nature our work is built upon probabilistic crash consistency: we skip cache-line flushing for those cache blocks with high probability of crash consistency. Chidambaram et al.~\cite{Chidambaram:sosp13} apply the idea of probabilistic crash consistency to journaling file systems. They ask
applications to order writes without incurring a disk flush, and request durability on disks when needed; they use checksums to detect data/metadata inconsistency. Our work is different from theirs in three perspectives: (1) We do not modify write orders; (2) Our checksum can correct inconsistency; (3) We highly optimize the performance of Archapt to minimize runtime overhead in NVM (as main memory), while some techniques (e.g., reuse after notification) in~\cite{Chidambaram:sosp13} can cause high overhead in our scenarios.


The lazy persistency~\cite{8416846} is 
another example of probabilistic crash consistency. 
It focuses on loop-based HPC applications and skips cache-line flushing for \textit{all} dirty data objects. It relies on periodical cache flushing of all dirty cache blocks. Our work is different from them in two perspectives. (1) The lazy persistency does not decide which cache-line flushing can be skipped. (2) The lazy persistency cannot correct data inconsistency after a crash. The above two limitations can cause unpredictable data loss in committed transactions. 



\vspace{-5pt}
\section{Conclusions}
\vspace{-5pt}
Enabling high performance transaction is critical to release the performance benefit of persistent memory for many applications.
In this paper, we present \name{}, an architecture-aware, high performance transaction runtime system for persistent memory. \name{} reduces the number of cache-line flushing to improve performance of transactions. \name{} estimates if cache blocks of a persistent objects are in the cache to determine the necessity of cache-line flushing. Relying on checksum mechanisms to detect and correct data inconsistence, \name{} provides strong crash consistency. \name{} also coalesces cache blocks with low dirtiness to improve the efficiency of cache-line flushing.
Our results show that on average, \name{} reduces cache flushing by 66\% and improves system throughput by 19\% (42\% at most), 
using a undo logging (PMDK) and a redo logging (Mnemosyne) as baseline.


\bibliographystyle{IEEEtran}
\bibliographystyle{plain}
\bibliography{kai,li}

\end{document}